\documentclass[floatfix,prx,twocolumn,letterpaper,lengthcheck,superscriptaddress,showpacs,amssymb,amsmath,amsfonts,aps,altaffilletter,nofootinbib,nopreprintnumbers,showpacs,longbibliography]{revtex4-1}

\usepackage{color}
\usepackage{hyperref}
\usepackage{amsmath}
\usepackage{multirow}
\usepackage{graphicx}
\usepackage{caption}
\usepackage{subcaption}
\usepackage{placeins}

\newcommand{\ttouch}{\ensuremath{T_{\rm touch}}}

\newcommand{\tmin}{\ensuremath{T_{\rm min}}}

\newcommand{\tAH}{\ensuremath{T_{\rm common}}} 

\newcommand{\Surf}{\ensuremath{\mathcal{S}}}
\def\insubscript{\rm inner}
\def\outsubscript{\rm outer}

\newcommand{\Sout}{\Surf_{\outsubscript}}

\newcommand{\Sin}{\Surf_{\insubscript}}

\newcommand{\Sone}{\Surf_{1}}

\newcommand{\Stwo}{\Surf_{2}}

\newcommand{\Munit}[1][\,]{\ensuremath{#1\mathcal{M}}}

\begin{document}

\title[]{Self-intersecting marginally outer trapped surfaces}

\author{Daniel Pook-Kolb} 
\affiliation{Max-Planck-Institut f\"ur Gravitationsphysik (Albert
  Einstein Institute), Callinstr. 38, 30167 Hannover, Germany}
\affiliation{Leibniz Universit\"at Hannover, 30167 Hannover, Germany}

\author{Ofek Birnholtz}
\affiliation{Center for Computational Relativity and Gravitation,
  Rochester Institute of Technology,
  170 Lomb Memorial Drive, Rochester, New York 14623, USA}

\author{Badri Krishnan} 
\affiliation{Max-Planck-Institut f\"ur Gravitationsphysik (Albert
  Einstein Institute), Callinstr. 38, 30167 Hannover, Germany}
\affiliation{Leibniz Universit\"at Hannover, 30167 Hannover, Germany}

\author{Erik Schnetter}
\affiliation{Perimeter Institute for Theoretical Physics, Waterloo, 
  ON N2L 2Y5, Canada}
\affiliation{Physics \& Astronomy Department, University of Waterloo,
  Waterloo, ON N2L 3G1, Canada}
\affiliation{Center for Computation \& Technology, Louisiana State
  University, Baton Rouge, LA 70803, USA}

\date{2019-06-28}

\begin{abstract}

  We have shown previously that a merger of marginally outer trapped
  surfaces (MOTSs) occurs in a binary black hole merger and that there
  is a continuous sequence of MOTSs which connects the initial two
  black holes to the final one.  In this paper, we confirm this
  scenario numerically and we detail further improvements in the
  numerical methods for locating MOTSs.  With these improvements, we
  confirm the merger scenario and demonstrate the existence of
  self-intersecting MOTSs formed in the immediate aftermath of the
  merger.  These results will allow us to track physical quantities
  across the non-linear merger process and to potentially infer
  properties of the merger from gravitational wave observations. 

\end{abstract}

\maketitle

\section{Introduction}
\label{sec:intro}

Numerous binary black hole merger events have now been observed by
gravitational wave detectors
\cite{Abbott:2016blz,TheLIGOScientific:2016pea,LIGOScientific:2018mvr,Green:2017voq,Zackay:2019tzo,Nitz:2018imz,Venumadhav:2019lyq}.
The general features of the gravitational wave signal from such events
are now well known.  The first is the inspiral regime where the signal
is a chirp of increasing amplitude and frequency, and the system is
effectively modeled as two point particles orbiting around each other
and emitting gravitational waves as the orbit decays.  As the two
black holes approach each other and coalesce to form a final common
black hole, the inspiral description is no longer valid, and
non-perturbative aspects of general relativity become important; this
is the merger regime.  Eventually, as the final black hole reaches
equilibrium, the gravitational wave signal can be well modeled as a
superposition of damped sinusoids (and, in principle, much weaker
power-law tails).  Corresponding to this behavior of the gravitational
wave signals, one visualizes the black holes themselves separately in
the three different regimes. The inspiral regime consists of two
disjoint black hole horizons slightly distorted by each other's
gravitational field.  The merger is visualized as two horizons very
close to each other and merging to form a single horizon which is
initially very distorted. Finally, the ringdown is modeled as a
perturbed Kerr horizon settling down to a final equilibrium Kerr black
hole.

These features of the waveform must be correlated in some way with
properties of the gravitational field in the strong field region.  In
particular, the three regimes must correspond in some way to
properties of the black hole horizons.  The details of the
correlations between the different portions of the gravitational wave
signal and the behavior of the horizons, and the precise demarcations
between the three regimes are yet to be fully quantified.  A full
understanding of these correlations is obviously necessary to have a
complete picture of a binary black hole merger (see
e.g. \cite{Gupta:2018znn,Jaramillo:2011rf,Jaramillo:2012rr,Kamaretsos:2012bs,Kamaretsos:2011um,Bhagwat:2017tkm}).
It is also of interest to understand further quantitative features of
the merger, such as the evolution of physical quantities across the
merger.  This includes, among other things, the fluxes of energy and
angular momentum, and the evolution of higher order multipoles during
the merger.  These might be correlated with interesting features of
the radiative multipoles found in \cite{Borhanian:2019kxt}.  Numerical
simulations are capable of solving the Einstein equations with high
accuracy for binary black hole mergers (see
e.g. \cite{Pretorius:2005gq,Campanelli:2005dd,Baker:2005vv,Szilagyi:2009qz}).
Such simulations provide an obvious avenue for exploring such
questions.

To understand the correlations between the gravitational wave signal
and the black hole horizons, we need to first decide precisely what
geometrical quantities on the horizon should be considered.  In fact,
we need to go a further step backwards and decide what kind of
horizons should be considered.  There are two different ways of
visualizing horizons using either event horizons or marginally trapped
surfaces.  Both of these descriptions are in good agreement in the
inspiral and ringdown regimes, but differ substantially during the
merger where non-linear and non-perturbative effects of general
relativity are especially important.  Consider first the event horizon
description.  An event horizon is the boundary of the region which is
causally disconnected from an asymptotically far away observer.  It is
clear that locating an event horizon requires knowledge of the global
properties of the spacetime infinitely far into the future.  It is
possible, though not trivial, to locate event horizons in numerical
binary black hole simulations
\cite{Hughes:1994ea,Diener:2003jc,PhysRevLett.74.630,Thornburg:2006zb},
and this yields the well known ``pair of pants'' picture
\cite{Matzner:1995ib}.  The cross-sections of the ``pair of pants''
corresponds with the expectations described above.  At early times,
the cross-section of the event horizon consist of two disjoint
surfaces corresponding to the two separate black holes, and a single
spherical surface at the end.  There are several interesting features
of the event horizon in the merger, including the existence of a
toroidal phase early in the merger and the non-differentiability of
the event horizon \cite{PhysRevD.60.084019}; the non-differentiability
is in fact a general feature of event horizons
\cite{Chrusciel:2000gj,Chrusciel:1996tw}.

The ``pair of pants'' picture is intuitively appealing and moreover it
seems to provide a complete picture of the black hole merger in
accordance with our physical expectations.  In reality however, this
picture is not so useful, both as a matter of principle and therefore
also for any detailed quantitative studies.  The problems can be
traced back to the global and teleological nature of event horizons:
to locate them, one needs to know what happens in the spacetime far in
the future.  In perturbative situations and when the end-state is
known or assumed, it is indeed possible to obtain expressions for the
fluxes of energy and angular momentum through the event horizon
\cite{Hawking:1972hy}.  In general dynamical situations however, this
is not true.  There are simple examples, even in spherical symmetry,
when the area of the event horizon grows without any corresponding
flux of energy \cite{Ashtekar:2004cn}.  Due to these teleological
properties, there is no possible local expression of general validity
for, say, the fluxes of energy and angular momentum through event
horizons.  It is thus not clear how to carry out the program of
understanding the merger and relating it to gravitational wave
observations outlined at the beginning of the previous paragraph.  As
a side remark, the teleological property also makes it difficult to
locate event horizons in numerical simulations in real time, but in
any case, it is certainly possible to locate them once the simulations
are complete.

There is an alternate way of visualizing a binary black hole merger
which, for both conceptual and practical reasons, is of much greater
importance in numerical simulations.  The starting point is an unusual
property of certain surfaces in the black hole region, first pointed
out by Penrose \cite{Penrose:1964wq}.  This requires the notion of the
expansion $\Theta$ of a congruence of light rays; $\Theta$ is the
logarithmic rate of change of an infinitesimal cross-section
transverse to the null geodesics.  A round sphere in flat space has
$\Theta>0$ for the outgoing light rays and $\Theta< 0$ for the ingoing
ones.  In the black hole region, there exist spheres (the trapped
surfaces) for which both sets of light rays have negative
expansion. The outermost such sphere at any given time has vanishing
outgoing expansion; these are the marginally trapped surfaces.  In
stationary situations such as for a Schwarzschild or Kerr black hole,
cross-sections of the event horizon are also marginally trapped
surfaces, but this correspondence is not true in non-stationary
situations.  Thus, cross-sections of the event horizon are marginally
trapped surfaces very early in the inspiral regime or at very late
times.  At intermediate times, especially near the merger, the two
notions are very different.  Furthermore, unlike event horizons,
marginal surfaces are not teleological and can be located at any given
time without reference to any future properties of spacetime.  It is
possible to define physical quantities such as mass, angular momentum,
multipole moments, and fluxes of energy and angular momentum
quasi-locally, i.e. on the marginal surfaces.  For this reason,
marginal surfaces are widely used in numerical simulations when
referring to the properties of black holes.  There is a large
literature on these quasi-local definitions and their applications to
various problems in classical and quantum black hole physics (see
\cite{Ashtekar:2004cn,Booth:2005qc,Faraoni:2015pmn,Krishnan:2013saa}
for reviews).

Despite this progress, there is still a missing ingredient, namely a
unified treatment of inspiral, merger and ringdown.  Thus far, all
studies of binary black hole coalescence using marginal surfaces have
considered the pre- and post-merger regimes separately.  The reason
for this is that, until recently, it was not known how marginal
surfaces behave across the merger; near the merger the marginal
surfaces are extremely distorted and previous numerical methods were
not successful in tracking such highly distorted surfaces.  Using
improved numerical methods \cite{Pook-Kolb:2018igu}, we have recently
shown the first evidence for the existence of a continuous sequence of
marginal surfaces which interpolates between the two disjoint initial
black holes and the single final remnant black hole
\cite{Pook-Kolb:2019iao}.  This is the analog of the ``pair of pants''
picture for event horizons.  In the present work, with further
improvements in numerical methods for locating marginal surfaces, we
shall provide further unambiguous evidence for this scenario.  We
shall also show the existence of marginal surfaces with
self-intersections.  In a companion paper we shall study physical
characteristics of the world-tube of marginal surfaces, which is the
other important ingredient for physical applications.

\begin{figure*}[h]
    \includegraphics[width=0.8\textwidth]{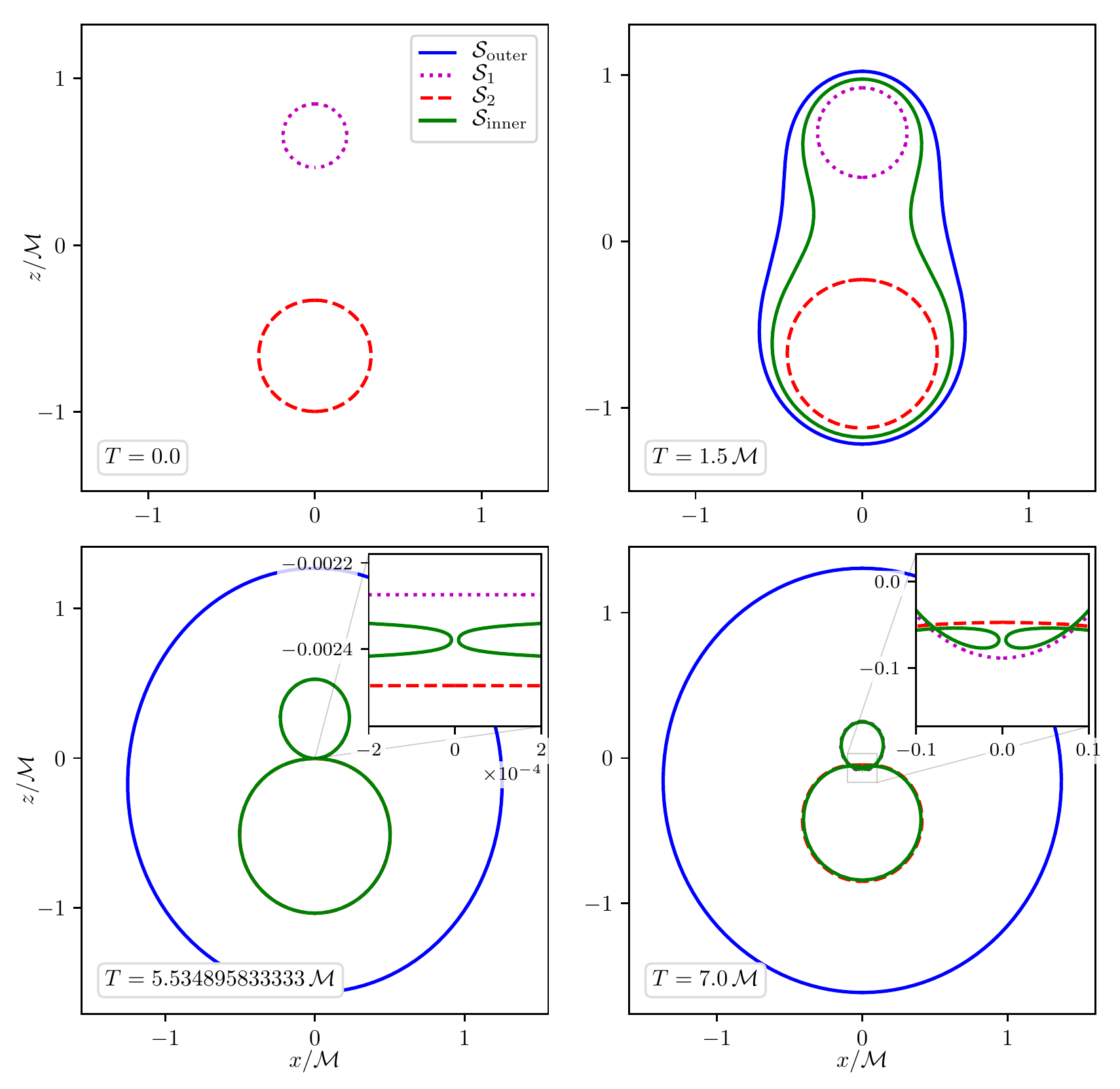}
    \caption{%
        MOTS structure of a simulation of Brill-Lindquist initial
        data shown at different simulation times.
        The self-intersection of $\Sin$ is present from the first
        instance it is found after $\Sone$ and $\Stwo$ touch at
        $\ttouch \approx 5.5378\Munit$.
        The upper left panel shows the initial condition and the upper
        right panel a time shortly after the two common MOTSs $\Sout$
        and $\Sin$ have formed together. The lower left panel shows the
        last time we were able to locate $\Sin$ before $\Sone$ and
        $\Stwo$ touch and then start to intersect, while the lower
        right panel shows a time well after $\ttouch$.
    }
    \label{fig:merger1}
\end{figure*}

The scenario we obtain for the merger is summarized in
Fig.~\ref{fig:merger1}. The details showing how these results are
obtained will be explained in the next sections. The figure shows four
snapshots of the MOTSs at various times\footnote{%
    We define the factor $\Munit[] := M_{\rm ADM} / 1.3$
    to be able to state our coordinate quantities in terms of the
    ADM mass, which in our simulations was chosen to be $1.3$.
}
in a head-on binary black hole
merger starting with Brill-Lindquist initial data.  We initially have
only the two individual MOTSs without a common horizon.  As the black
holes get closer, a common MOTS is formed which immediately bifurcates
into outer and inner portions visible in the second snapshot.  The
outer portion loses its distortions as it approaches its equilibrium
state, while the inner MOTS becomes increasingly distorted.  At some
point, just shortly after the third snapshot, the two individual MOTSs
touch each other exactly at the time when they merge with the inner
common MOTS.  After this merger, the two individual MOTSs go through
each other. Surprisingly, it turns out that the inner common MOTS
continues to exist after the merger and now has self-intersections as
shown in the last snapshot.  The remainder of this paper will be
devoted to explaining how we arrive at this result.  A detailed study
of the physical aspects of this scenario will be presented elsewhere.

Sec.~\ref{sec:motsdefn} summarizes the basic definitions and concepts
that we shall need for this paper.  The improved numerical algorithm
for locating marginal surfaces is described in
Sec.~\ref{sec:motsfinder} and Sec.~\ref{sec:validation} shows various
numerical tests to validate the method.  Sec.~\ref{sec:numerics}
discusses our modifications to the numerical methods used to evolve
Cauchy data using the Einstein equations. These modifications allow us
to reach the required numerical accuracy and convergence, and to carry
out our simulations more efficiently.  Sec.~\ref{sec:selfintersect}
puts together all these ingredients and presents our main results.
For a particular initial configuration (the head on collision of
comparable mass non-spinning black holes), the merger of marginally
trapped surfaces is demonstrated with high numerical accuracy.  The
merger involves the formation of a marginally trapped surface with
self-intersections, showing topology change in a binary black hole
merger.

\section{Marginally outer trapped surfaces}
\label{sec:motsdefn}

Let $\ell^a$ be a congruence of future directed null geodesics, and
let $n^a$ be another such congruence satisfying $\ell^a n_a = -1$.
Let $q_{ab}$ be the Riemannian metric in the 2-dimensional space
transverse to both $\ell^a$ and $n^a$.  The divergence of $\ell^a$ and
$n^a$ are respectively
\begin{equation}
  \Theta_{(\ell)} = q^{ab}\nabla_a\ell_b\,,\quad \Theta_{(n)} = q^{ab}\nabla_an_b\,.
\end{equation}
Let $\Surf$ be a closed spacelike 2-surface with null normal fields
$\ell^a$ and $n^a$ respectively.  We assume that it is possible to
assign outgoing and ingoing directions on $\Surf$, and by convention,
$\ell^a$ and $n^a$ are the outgoing and ingoing null normals
respectively.  The classification of $\Surf$ based on conditions on the
expansions are the following:
\begin{itemize}
\item Trapped: $\Theta_{(n)}< 0$, $\Theta_{(\ell)} < 0$
\item Un-trapped: $\Theta_{(n)}< 0$, $\Theta_{(\ell)} > 0$
\item Marginally trapped: $\Theta_{(n)}< 0$, $\Theta_{(\ell)} = 0$
\item Marginally outer-trapped: $\Theta_{(\ell)} = 0$ (no condition on
  $\Theta_{(n)}$)
\end{itemize}
All of these refer to future-directed $\ell^a$. Thus we should say
future-trapped rather than just trapped, but we shall only consider
future directed cases.  The most important case for us is the
marginally outer trapped surface (MOTS) lying within a spatial slice $\Sigma$.

As mentioned in the introduction, there is a large literature on the
application of MOTSs to study black holes in various contexts (see
e.g. \cite{Ashtekar:2004cn,Booth:2005qc,Gourgoulhon:2005ng,Hayward:2004fz,Jaramillo:2011zw,Krishnan:2013saa,Krishnan:2007va}). They
are regularly used in numerical relativity simulations to compute
physical quantities
\cite{Dreyer:2002mx,Schnetter:2006yt,Gupta:2018znn}, and this
formalism leads naturally to various versions of quasi-local black
hole horizons.  

While we shall not delve into the mathematical and physical
characteristics of MOTSs here, it shall be useful to understand the
stability operator for a MOTS and its relevance for time evolution.
For a given MOTS $\Surf$ consider a smooth one-parameter family of
closed spherical surfaces $\Surf_\lambda$ which are \emph{variations}
of $\Surf$ in the normal direction \cite{Newman1987} within the
spatial hypersurface $\Sigma$.

On each $\Surf_\lambda$, just as
for $\Surf$, we can define the null normals and calculate the expansion
$\Theta_{(\ell)}(\lambda)$, which will of course generally not
vanish. The differentiation of $\Theta_{(\ell)}(\lambda)$ leads to an
operator $L$ on $\Surf$:
\begin{equation}
  \delta_{fr}\Theta_{(\ell)} =: Lf\,.
\end{equation}
Here $r^a$ refers to the unit outward pointing spacelike normal to
$\Surf$ (within $\Sigma$) and $f$ is a scalar function on $\Surf$.
Along the 1-parameter family $\Surf_\lambda$, every point on $\Surf$
traces out a curve with tangent vector $fr^a$.  The variation of the
expansion, i.e. the left hand side of the above equation, is the
derivative of the expansion along these curves.  This procedure
defines an elliptic operator $L$ on a MOTS and the precise expression
for $L$ can be worked out.  Generically it is of the form
\begin{equation}
\label{eq:stability_operator}
  Lf = -\Delta f + \gamma^a \partial_a f + \beta f\,,
\end{equation}
Here $\Delta$ is the Laplace-Beltrami operator on $\Surf$ compatible
with $q_{ab}$, $\gamma^a$ is a vector field on $\mathcal{S}$ related
to black hole spin, and $\beta$ is a scalar related to the intrinsic
(two-dimensional) Ricci scalar of $\Surf$.  Thus, $L$ is not
necessarily a self-adjoint operator due to the presence of $\gamma^a$,
and its eigenvalues are not necessarily real.  Nevertheless, its
principal eigenvalue $\Lambda_0$, i.e. the eigenvalue with the
smallest real part is indeed real.  In this paper we shall restrict
ourselves to non-spinning black holes with vanishing $\gamma^a$ so
that all eigenvalues are real.

The primary utility of $L$ is that it determines the behavior of
$\Surf$ under time evolution.  It was shown that if the principal
eigenvalue is positive, then the MOTS evolves smoothly in time
\cite{Andersson:2005gq,Andersson:2007fh,Andersson:2008up}. This
stability condition is equivalent to saying that an outward
deformation of $\mathcal{S}$ makes it untrapped which is what we
expect to happen for the apparent horizon.  While not emphasized in
\cite{Andersson:2005gq,Andersson:2007fh,Andersson:2008up}, the
condition for the existence of $\Surf$ under time evolution is the
invertibility of $L$. Thus, if 0 is not in the spectrum of $L$, then
$\Surf$ continues to evolve smoothly.  In the case when $\Lambda_0<0$
(which will happen in our case), we must consider the next eigenvalue
$\Lambda_1$ and check that it does not vanish.  See
e.g. \cite{Booth:2017fob,Sherif:2018scu,Mach:2017peu} as examples of
studies which consider this notion of stability in specific examples.

\section{Numerical methods for locating highly distorted MOTSs}
\label{sec:motsfinder}

Consider a Cauchy surface $\Sigma$ on which we wish to locate a MOTS
$\Surf$.  Let $\Sigma$ be equipped with a Riemannian metric $h_{ij}$ with
the associated Levi-Civita connection $D_a$, and let the extrinsic
curvature of $\Sigma$ be $K_{ij}$. Let $r^a$ be the unit-spacelike
normal to $\Surf$ within $\Sigma$ and let $\tau^a$ be the unit-timelike
normal to $\Sigma$.  Then, a suitable choice of null-normals to $\Surf$ is
\begin{equation}
  \ell^a = \frac{1}{\sqrt{2}} \left(\tau^a+ r^a\right)\,,\quad n^a = \frac{1}{\sqrt{2}}\left(\tau^a - r^a\right)\,.
\end{equation}
The condition $\Theta_{(\ell)}=0$ is rewritten as
\begin{equation}
\label{eq:motsequation}
  D_ar^a + K_{ab}r^ar^b - K = 0\,.
\end{equation}
This is the equation that we must solve to find $\Surf$. The conventional
approach \cite{Thornburg:2006zb,Thornburg:2003sf} assumes that the
surface is defined by a level-set function
\begin{equation}
  \label{eq:starshaped}
  F(r,\theta,\phi) = r - h(\theta,\phi)\,,
\end{equation}
where $(r,\theta,\phi)$ are spherical coordinates on $\Sigma$.  This
assumes that $\Surf$ is \emph{star-shaped} with respect to the origin in
the chosen coordinate system.  In other words, any ray drawn from the
origin must intersect the surface only once.  This assumption will not
hold for the surfaces of interest for us. A variant of this method was
proposed in \cite{Pook-Kolb:2018igu} and shown to be capable of
locating extremely distorted surfaces.  This new method is based on
using a \emph{reference} surface $\sigma_R$, and representing $\Surf$
in terms of distances $h(\lambda, \mu)$ from $\sigma_R$, where
$\lambda, \mu$ parameterize $\sigma_R$.  As long as the reference
surface is chosen appropriately, the method can be used to locate
almost arbitrarily distorted surfaces.  For example, in a numerical
evolution, one could choose $\sigma_R$ to be the MOTS located in the
previous time step.
The problem of locating $\Surf$ then translates to solving a nonlinear
partial differential equation for the horizon function $h$.
This can be done e.g. via a pseudospectral method,
which is what we chose.

For our present application, we have implemented two additional
features compared to what was used in \cite{Pook-Kolb:2018igu}.  These
features are meant to deal with two additional complications that we
must necessarily deal with: i) surfaces which have a very narrow
``neck'' (almost like a figure-eight), and in some instances have
features like cusps and self intersections.  For this purpose,
motivated by the methods used in \cite{Jaramillo:2009zz}, we employ
bi-spherical coordinates \cite{Ansorg:2005bp}.
ii) Unlike in \cite{Pook-Kolb:2018igu} where
the MOTS finder was applied to analytical initial data, we now have to
deal with numerically generated data on a finite mesh.  This requires
the use of interpolation schemes some of which were already used in
\cite{Pook-Kolb:2019iao}.  We now describe in turn both of these
additional features.  We shall still be restricted to axisymmetry in
this work, reducing the task of finding the horizon function $h$ to
a one-dimensional problem.  However, no in-principle difficulties are
foreseen for general non-axisymmetric cases.

\subsection{Bi-spherical coordinates}
\label{subsec:bipolar}

\begin{figure}[h]
    \centering
    \includegraphics[width=0.48\textwidth]{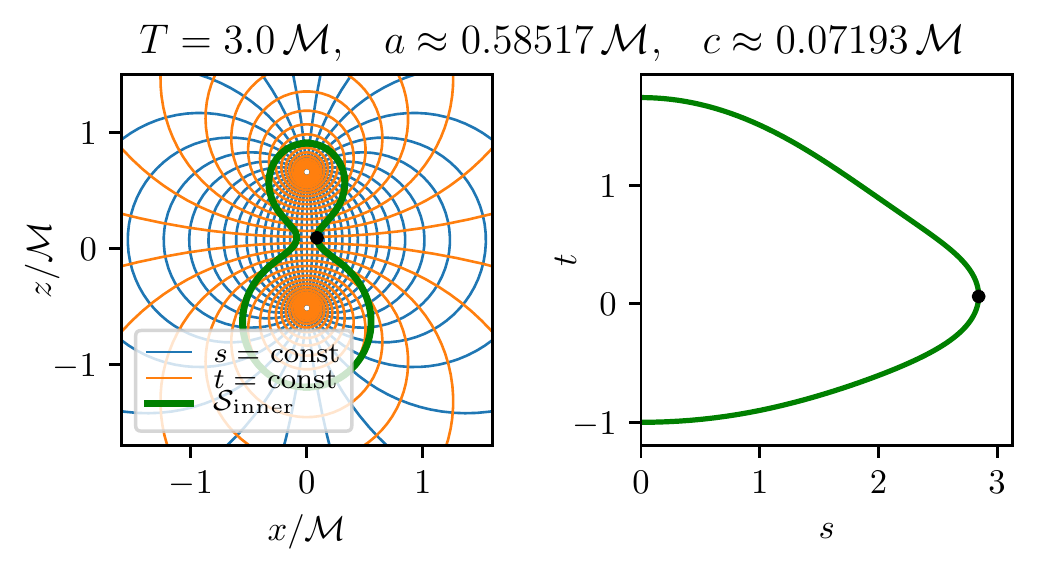}
    \includegraphics[width=0.48\textwidth]{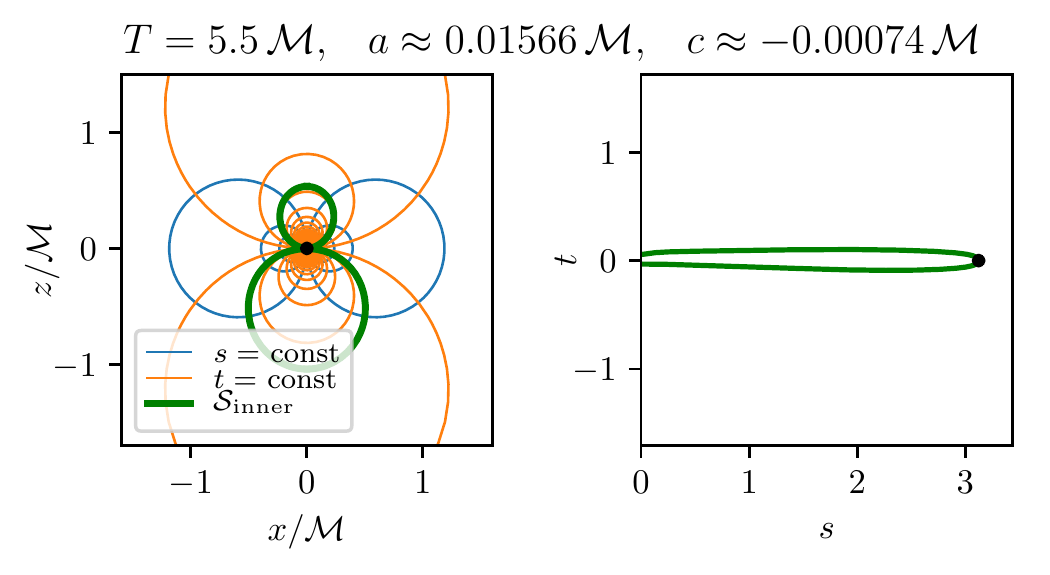}
    \includegraphics[width=0.48\textwidth]{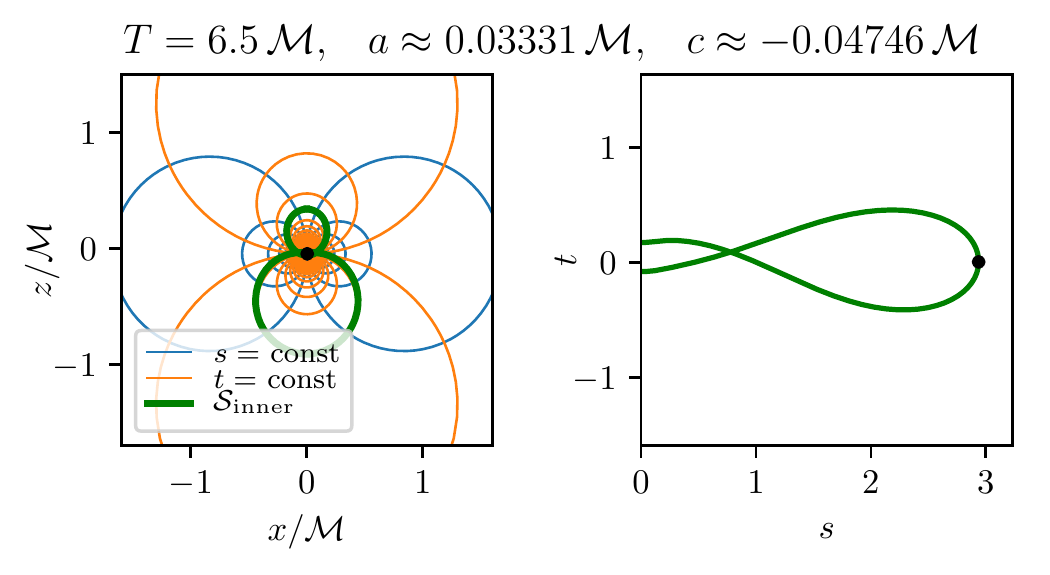}
    \caption{%
        Visualizations of $\Sin$ in bipolar coordinates at different
        simulation times $T$. The left column shows the MOTS and lines
        of constant $s$ and $t$ in the $(x,z)$ plane while the right
        column contains $\Sin$ in the $(t,s)$ plane. Note that only
        positive values of $s$ are shown, though the full MOTS is of
        course symmetric about $s=0$.
        The first row shows a slightly distorted MOTS in both
        representations. At $T=5.5\Munit$ (second row), $\Sin$ is highly
        distorted in the $(x,z)$ plane and only slightly distorted in
        the bi-spherical coordinates.
        The last row shows a case of a self-intersecting $\Sin$.
        The dot marks the location of the ``neck'' in all cases.
    }
    \label{fig:bipolar1}
\end{figure}
For axisymmetric surfaces, choosing the symmetry axis to be the
$z$ axis, we can restrict ourselves to the $(x,z)$ plane and it is
often convenient to characterize any point using polar coordinates,
i.e. using the distance from the origin and the angle of the position
vector with the $z$ axis.  However these coordinates are not optimal
for describing surfaces with a very narrow neck connecting two
spherical portions, i.e. close to a figure-eight in shape.  We use
instead the bipolar coordinates $(s,t)$ which are based on two foci
located at
$x=0$, $z=c \pm a$:
\begin{equation}\label{eq:bipolar}
  x = \frac{a\sin s}{\cosh t - \cos s}\,,\quad
  z = \frac{a\sinh t}{\cosh t - \cos s} + c\,.
\end{equation}
The $(s,t)$ coordinates make the highly distorted inner common MOTS
$\Sin$ much easier to parameterize.

Examples demonstrating the effect of this coordinate transformation
for three different simulation times are shown in
Fig.~\ref{fig:bipolar1}.  The three snapshots are at times i) $T=3\Munit$
which is a bit after the top right panel of Fig.~\ref{fig:merger1} and
$\Sin$ does not have extreme distortions; ii) $T=5.5\Munit$, shortly before
the bottom left panel in Fig.~\ref{fig:merger1} where $\Sin$ has a
very narrow neck, and finally iii) $T=6.5\Munit$, a little bit before the
bottom right panel of Fig.~\ref{fig:merger1}, and $\Sin$ has
self-intersections.

The bi-spherical coordinates are employed only for $\Sin$; none of the
other horizons have the narrow neck and these coordinates are
unnecessary to locate them.  To determine the value of $c$ in
\eqref{eq:bipolar}, we first find the two individual MOTSs $\Sone$ and
$\Stwo$ and choose $c$ to lie in the coordinate center between the
lowest point of $\Sone$ and the upper-most point of $\Stwo$.  As
detailed below, we find the various MOTSs in a series of time slices
produced by the numerical simulation.  During this {\em tracking} of
$\Sin$, we numerically approximate the optimal value for $a$ as a
post-processing step once the MOTS is located.  In practice, this is
done by representing $\Sin$ in bi-spherical coordinates and expressing
the coordinate functions $s(\lambda), t(\lambda)$ as a truncated
series of sines and cosines, respectively, which have the correct
symmetry for the problem. We use a slightly lower number of basis
functions than necessary to obtain convergence and check the residual
expansion of the now imperfect representation. Varying the parameter
$a$, we repeat this process to find the value resulting in the lowest
residual. The value for $a$ determined this way is then used for
finding the MOTS in the next slice, assuming the optimal parameter
varies slowly with simulation time.

A further optimization is to re-parameterize the reference surface
$\sigma_R$ prior to finding the MOTS. A natural choice of
parameterization would be the proper length or proper length in
coordinate space, the latter obviously being better suited for our
numerical representation of the surface. If the curve representing
$\sigma_R$ in coordinate space is
$\lambda \mapsto \gamma_R(\lambda)$, this would mean that
$\Vert\gamma_R'(\lambda)\Vert_2 \equiv \text{const}$.
However, we obtained faster convergence by taking a non-constant speed
function such that $\Vert\gamma_R'(\lambda)\Vert_2$ is
roughly\footnote{%
    We smoothen the speed function along the MOTS by exponentially
    damping the coefficients of a cosine series representation.
    This reduces higher frequencies in the density of collocation
    points along $\Surf$.
}
proportional to $1/k_{AB}k^{AB}$, where $k_{AB}$ is the second
fundamental form of $\sigma_R$ embedded in coordinate space.

Utilization of bi-spherical coordinates together with the above
re-parameterization has led to convergent solutions $\Sin$ with about
one order of magnitude fewer collocation points compared to the
previous method.

\subsection{Interpolating numerical data} 
\label{subsec:interpolation} 

In each time step, our axisymmetric numerical simulations produce
data on a 2-dimensional grid of points lying equidistant in the
$(x,z)$ coordinate plane. However, for the nonlinear search for
a MOTS $\Surf$, the expansion $\Theta_{(\ell)}$ and its derivatives
have to be computed on a set of points
$x_n\in\mathbb{R}^2$ along trial surfaces $\Surf^{i}$, c.f.
\cite{Pook-Kolb:2018igu}, Section III.B.
This requires evaluating the components of the metric $h_{ij}$, its
first and second spatial derivatives, the extrinsic curvature
$K_{ij}$ and its first spatial derivatives at the points $x_n$ which
generally do not coincide with any of the grid points of the
simulation.

In \cite{Pook-Kolb:2019iao} we used $4$th order accurate $5$-point
Lagrange interpolation. Derivatives were obtained by evaluating $4$th
order accurate finite differencing derivatives using the data on the
grid and then interpolating the results using $5$-point Lagrange
interpolation.
For the present paper, however, we switched to quintic Hermite
interpolation, which allows us to control the values along with first
and second derivatives of the interpolant at the grid points. These
derivatives are evaluated using $6$th order accurate finite
differencing.
Derivatives between the grid points are then computed by analytically
differentiating the interpolating polynomial. The advantage is that
now first and second derivatives are continuous throughout, which is
not the case with Lagrange interpolation.

Interpolation of discrete data will be more accurate with increased
grid resolution.  However, it will never be exact and even floating
point accuracy cannot be neglected, especially near the punctures at
computationally feasible resolutions.  These additional inaccuracies
may limit the numerical convergence as they move the plateau we see
below in Fig.~\ref{fig:spectral_convergence} up---for example when
moving closer to the punctures or reducing the grid resolution---or
down.  To account for this effect while tracking a MOTS through
simulation time, we compute the expansion between the collocation
points each time the expansion drops below a pre-set tolerance
{\em at} the collocation points.  After this, we increase the spectral
resolution and continue until the tolerance is met at the now larger
set of collocation points.  This is repeated until the expansion
between the collocation points no longer improves, signaling that we
have reached the plateau.

A second criterion for stopping to increase the spectral resolution is
derived from the absolute values of the coefficients $a_n$ of the
spectral representation of the horizon function $h$.
In a pseudospectral method using a basis of cosines, one expects these
coefficients to fall off exponentially for large $n$ if the solution
exists.  We hence stop increasing the resolution if sub-exponential
fall-off of the $a_n$ is found following a region of exponential
convergence.  This prevents our code from overfitting $\Surf$ to
features introduced by the interpolation method, which happens
especially for lower resolution simulations.

\section{Validating the MOTS finder}
\label{sec:validation}

With the addition of numerical simulations, the task for our MOTS
finder has become more general compared to the purely time-symmetric
cases considered in \cite{Pook-Kolb:2018igu}.  Therefore, and in light
of the surprising result of a self-intersecting MOTS, it is important
to validate the method and test it for correctness in an analytic case
where the result is known.  We shall later present convergence results
for further validation.

For this purpose we construct a non-time-symmetric slice with
analytically known horizon shape. We choose a slice of the
Schwarzschild spacetime in Kerr-Schild coordinates
\cite{Matzner:1998pt}, i.e.
\begin{eqnarray}\label{eq:schwarzschildKS}
    h_{ij} &=& \delta_{ij} + \frac{2m}{r} \frac{x_i x_j}{r^2}\,,\\
    \label{eq:schwarzschildKS_curv}
    K_{ij} &=& \frac{2m}{r^4} \frac{1}{\sqrt{1+2m/r}}
    \left[ r^2 \delta_{ij} - \left( 2 + \frac{m}{r} \right) x_i x_j \right]
    \,,
\end{eqnarray}
where $\delta_{ij}$ is the flat metric, $x_i$ are the standard
Cartesian coordinates for the flat metric, and we shall often use
$(x,y,z)$ instead of $x_i$ when no confusion can arise.  For
Schwarzschild, the radial coordinate is just $r^2= x^2+y^2+z^2$.
These data have nontrivial extrinsic curvature with the horizon being
located at $r=2m$.

\begin{figure}[h]
    \centering
    \includegraphics[width=0.4\textwidth]{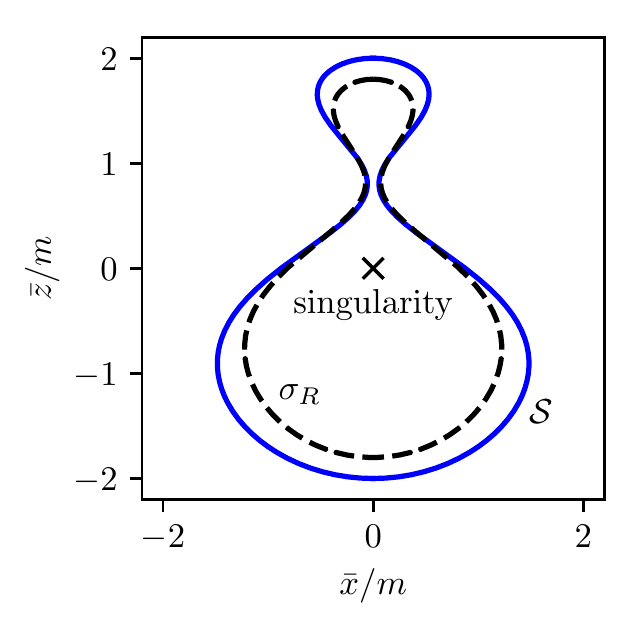}
    \includegraphics[width=0.45\textwidth]{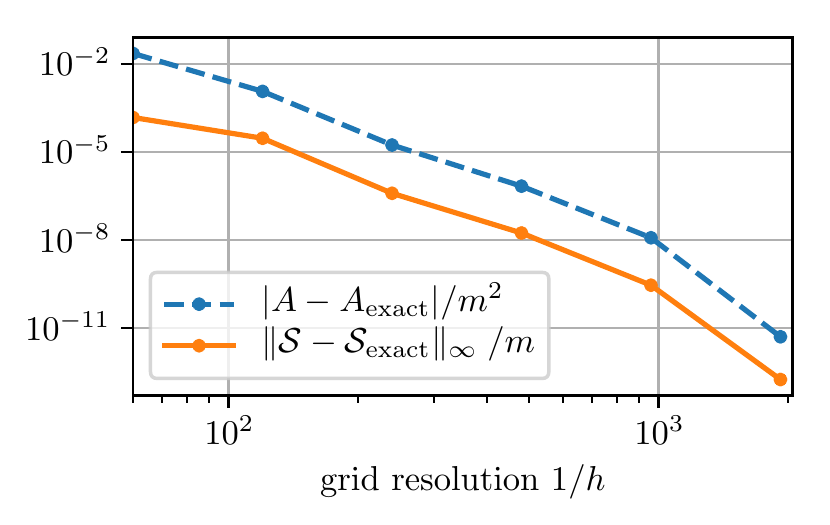}
    \caption{%
        {\em Top}: Horizon $\Surf$ and reference shape $\sigma_R$ for
        the transformed slice of Schwarzschild spacetime. The
        parameters for the transformation via
        \eqref{eq:schwarzschildCoordTransform} are
        $\beta = 0.97$, $\gamma = 0.7\Munit$ and $z_0 = 0.8\Munit$.
        {\em Bottom}: Convergence of the area (dashed) and surface
        coordinate shape (solid) with increased grid resolution.
        In each case, the spectral resolution was chosen such that a
        further increase does not result in a lower residual
        expansion (see section~\ref{subsec:convergence}).
        This thus shows the error introduced by the spatial
        discretization and interpolation.
    }
    \label{fig:schwarzschildKW-plot}
\end{figure}

To make the horizon non-star-shaped and thus the task more difficult
(but still axisymmetric), we transform the coordinates
$(x,z) \rightarrow (\bar{x},\bar{z})$ via
\begin{eqnarray}\label{eq:schwarzschildCoordTransform}
    \bar x = x \left(1 - \frac{\beta}{\cosh((z-z_0)/\gamma)}\right)
    \,,\quad
    \bar z = z\,.
\end{eqnarray}
These equations are used to sample $h_{ij}$ and $K_{ij}$ on grids of
various resolutions from $1/h = 30$ to $1/h = 1920$. We choose a
reference shape that is close but not identical to the horizon.
The MOTS $\Surf$ and the reference shape $\sigma_R$ are shown in the
first panel of Fig.~\ref{fig:schwarzschildKW-plot}.
For this test we compute the area $A$ of $\Surf$ and compare it to
the exact area $A_{\rm exact} = 16\pi m^2$, where $m=1$. We also
compute the maximum coordinate distance
$\Vert\Surf-\Surf_{\rm exact}\Vert_\infty$ of the numerical solutions
to the exact horizon.
The second panel demonstrates that our numerical solutions converge to
the expected solutions as the resolution of the numerical grid is
increased.

\section{The numerical evolutions}
\label{sec:numerics}

\subsection{Formulations, Discretization, and Implementation}

We set up initial conditions for the spacetime geometry as two
puncture black hole using the method of Brill and Lindquist
\cite{PhysRev.131.471}.
To evolve the geometry, we use the BSSN formulation of the Einstein
equations
with a $1+\log$ slicing and a $\Gamma$-driver shift condition
\cite{Alcubierre:2000xu, Alcubierre:2002kk}. We also impose
axisymmetry throughout the calculation.

For our setup (see below), we choose a domain with $x \in [0;10]$,
$z \in [-10;10]$, and $T \in [0;7]$. (Due to axisymmetry, we only
consider the hyperplane $y = 0$.) For simplicity, we use Dirichlet
boundary conditions to set all time derivatives to zero at the outer
boundary.
We check that the errors introduced by the artificial boundary
conditions do not affect the geometry near the MOTSs.

We choose a Cartesian basis for the tangent space, i.e. we represent
vectors and tensors via their $x, y, z$ components. Although
axisymmetry requires that certain components or linear
combinations of components must vanish, we do not explicitly impose
such conditions. Instead, we only impose axisymmetry on spatial
derivatives: We require that the Lie derivatives of all quantities in
the $\phi$ direction be zero, and we use this to remove all $y$
derivatives. ($y$ derivatives are then either $0$, or are replaced by
combinations of various $x$ derivatives.) We use l'H\^opital's rule
to regularize these expressions on the axis. This closely follows the
approach described in \cite{Pretorius_2005}, extended to handle second
derivatives as well. The set of expressions for handling first and
second $y$ derivatives for all tensor ranks appearing in the BSSN
formulation is lengthy, and is available in a Mathematica script as
part of Kranc \cite{Husa:2004ip, Kranc:web}.

In our discretization, we also require a small region ``on the other
side'' of the axis (where $x<0$), which we calculate by rotating the
region with $x>0$ by $\pi$.

We also experimented with the \emph{Cartoon} method
\cite{Alcubierre:1999ab} to impose axisymmetry. Cartoon uses a spatial
rotation in the $\phi$ direction and then spatial interpolation to
populate points away from the $y$ axis, so that $y$ derivatives can be
calculated in the standard manner. We found that the Cartoon method
does not work well with higher order (higher than $4$th) finite
differencing: The result of a Lagrange interpolation is not
continuous, which leads to large oscillations when derivatives are
taken near the axis where the Cartoon rotation angle is large.

In our setup, the punctures are located on the $z$ axis and are
initially at $z_\pm = \pm 0.65$. The puncture masses are $m_+ = 0.5$
and $m_- = 0.8$ (i.e. the ``upper'' black hole is smaller). The
punctures have no linear or angular momentum.

Details of initial and gauge conditions are described in
\cite{wardell_barry_2016_155394}. Our exact parameter settings are
available in the parameter files in the repository
\cite{pook_kolb_daniel_2019_3260171}.

We use $6$th order finite differencing to discretize space. We also
add a $6$th order Kreiss-Oliger artificial dissipation, which reduces
our spatial accuracy to $5$th order. We use a $6$th order accurate
Runge-Kutta time integrator. Our discretization is globally $5$th
order accurate, as we demonstrate below in section
\ref{sec:convergence}. We do not use
mesh refinement nor multiple grid patches as these would not be
beneficial for our calculations that span only a short time
and a small region of space, compared
to systems of orbiting binary black holes.

Compared to $4$th and $8$th order discretizations, $6$th order is most
efficient for us. $4$th order calculations require significantly
higher resolutions, and $8$th order calculations are significantly
slower since they use larger stencils and require more integrator
substeps. $8th$ order calculations also require higher resolutions
before their error falls below that of $6$th order calculations.

We perform our calculation via the \emph{Einstein Toolkit}
\cite{Loffler:2011ay, EinsteinToolkit:web}. We use \emph{TwoPunctures}
\cite{Ansorg:2004ds} to set up initial conditions and an axisymmetric
version of \emph{McLachlan} \cite{Brown:2008sb} to solve the Einstein
equations, which uses \emph{Kranc} \cite{Husa:2004ip, Kranc:web} to
generate efficient C++ code.

\subsection{Accuracy, Convergence}
\label{sec:convergence}

\begin{figure}[h]
    \centering
    \includegraphics[width=0.48\textwidth]{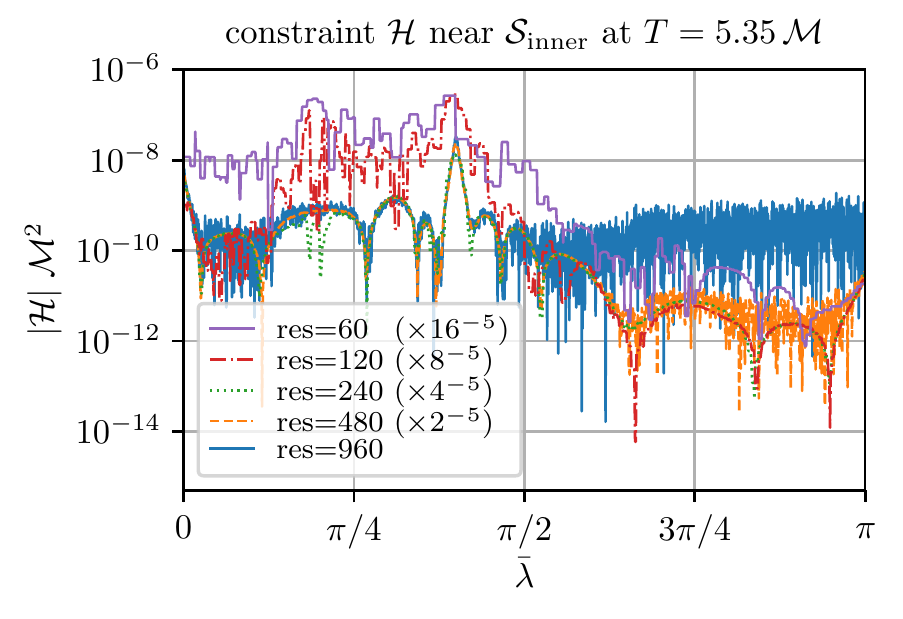}
    \includegraphics[width=0.48\textwidth]{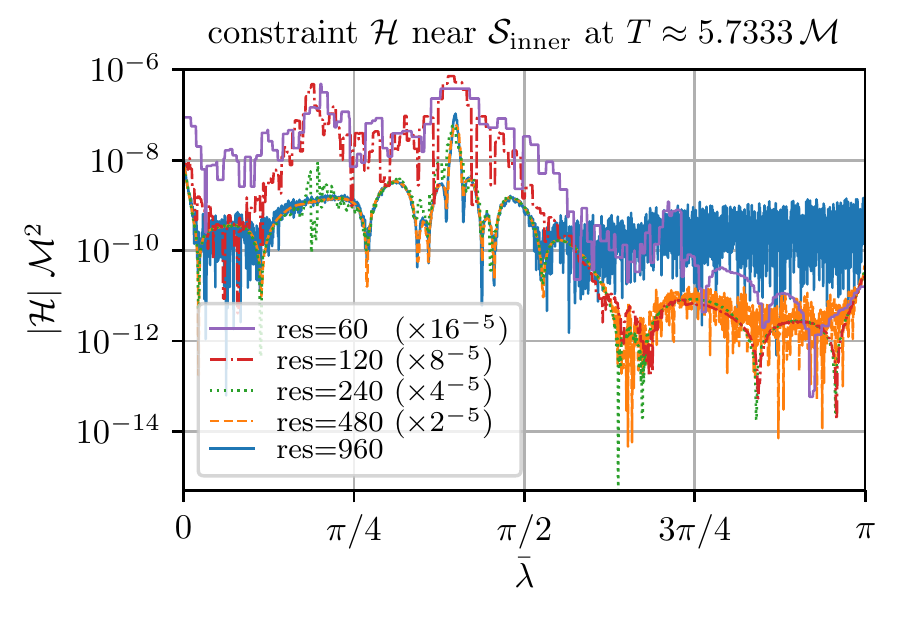}
    \caption{%
        Convergence of the Hamiltonian constraint
        for increasing resolutions  $1/h = 60$, $120$, $240$, $480$,
        $960$ at one time step before (upper panel) and after (lower
        panel) the individual horizons touch.
        The constraint is computed at grid points close to $\Sin$ and
        plotted over the proper length (scaled to $[0,\pi]$) of
        the curve representing $\Sin$ in the $(x,z)$ plane.
    }
    \label{fig:constr}
\end{figure}

To demonstrate the accuracy of our discretization,
we plot in Fig.~\ref{fig:constr}
the Hamiltonian constraint
\begin{equation}\label{eq:ham}
    \mathcal{H} = K_{ab} K^{ab} - K - R
\end{equation}
on grid points close to the inner common MOTS at two different times
for different grid resolutions.
Here, $R$ is the Ricci scalar of the slice $\Sigma$.
There is no significant difference between the two times.  Note that in
coordinate space, $\Sin$ lies closer to the punctures in its upper than
in its lower half, compare also Fig.~\ref{fig:merger1}.
In terms of the curve's proper length parameter $\bar\lambda$ (scaled
to $\bar\lambda\in[0,\pi]$), this
corresponds to $\bar\lambda \lesssim \pi/2$ and
$\bar\lambda \gtrsim \pi/2$, respectively, where our representation
only covers half of the plotted MOTS (say for positive $x$
values) due to axisymmetry.

The results have been scaled to account for $5$th order convergence.
We indeed find $5$th order convergence for $1/h \geq 240$
closer to the punctures and for $1/h \geq 120$ further away from
the punctures.  In that latter region, the highest resolution results
with $1/h = 960$ show slightly larger errors than expected from
$5$th order accuracy.

\begin{figure}[h]
    \centering
    \includegraphics[width=0.48\textwidth]{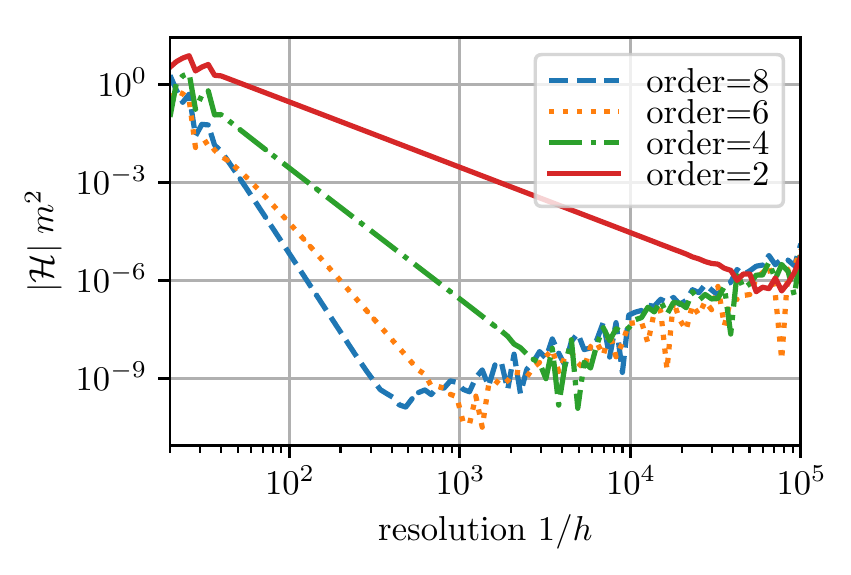}
    \caption{%
        Hamiltonian constraint computed at one point of a
        slice of the Schwarzschild spacetime in Kerr-Schild
        coordinates as defined in
        \eqref{eq:schwarzschildKS}, \eqref{eq:schwarzschildKS_curv}
        for grid resolutions $1/h = 20$ to $1/h = 10^5$.
        Since this is an exact solution of the Einstein equations, we
        expect $\mathcal{H} \equiv 0$,
        and this figure thus shows the discretization error.
        The constraint is evaluated at a coordinate distance of
        $r \approx 0.24\,m$ from the puncture.
    }
    \label{fig:differentiation_roundoff}
\end{figure}

This is caused by round-off errors starting to dominate the finite
difference derivatives, as is demonstrated in
Fig.~\ref{fig:differentiation_roundoff}.
Here, the different curves represent the results obtained using
stencils of $3$ to $9$ points for the derivatives of the metric
components, corresponding to $2$nd to $8$th order accuracy.  We see
the typical behavior of convergence up to the resolution at which the
round-off error becomes dominant.  This happens at lower resolutions
for the higher order methods as these reach the round-off limit
earlier.  Note that the optimal resolution depends on the function
being approximated and in our case becomes larger the closer we get to
the puncture.  This explains the different behavior in the first and
second half of the plots in Fig.~\ref{fig:constr}.

\section{The existence of self-intersecting MOTSs}
\label{sec:selfintersect}

With the technical improvements at hand, we now turn to the main
result of this paper, namely the merger of the inner MOTS with the two
individual horizons, and the occurrence of self intersecting MOTSs
just after this merger (see Fig.~\ref{fig:merger1}).  We will study a
single configuration with high resolution.  We focus primarily on
numerical accuracy and convergence to confirm the merger scenario and
the existence of self-intersecting MOTSs. There are obviously numerous
physical and geometrical properties of great interest.  First however,
we need to prove this scenario numerically beyond any reasonable
doubt, which is what we shall do here.  A detailed discussion of the
interesting physical and geometrical properties of the world tube of
MOTSs will be postponed to a forthcoming paper.  Similarly, we shall
not discuss here the various extensions to non-time symmetric and
non-axisymmetric data.  As mentioned previously, we start with
Brill-Lindquist initial data with the bare masses $m_+ = 0.5$ and
$m_- = 0.8$.  The initial coordinate separation between the punctures
is $1.3\Munit$ (i.e. $1$ in units of the total ADM mass
$M_{\rm ADM} = m_+ + m_-$).
Simulations are performed at
various grid resolutions: $1/h = 60$, $120$, $240$, $480$, $960$. We
have already shown in the previous section that the numerical solution
of the Einstein equations for the given initial data is sufficiently
accurate and all constraint violations converge at the expected rate
when $h$ is varied.  Given this numerical spacetime, we can use our
horizon finder to locate the various MOTSs.  It remains to be shown
now that the surfaces thus found are indeed MOTSs.

\begin{figure}[h]
    \centering
    \includegraphics[width=0.5\columnwidth]{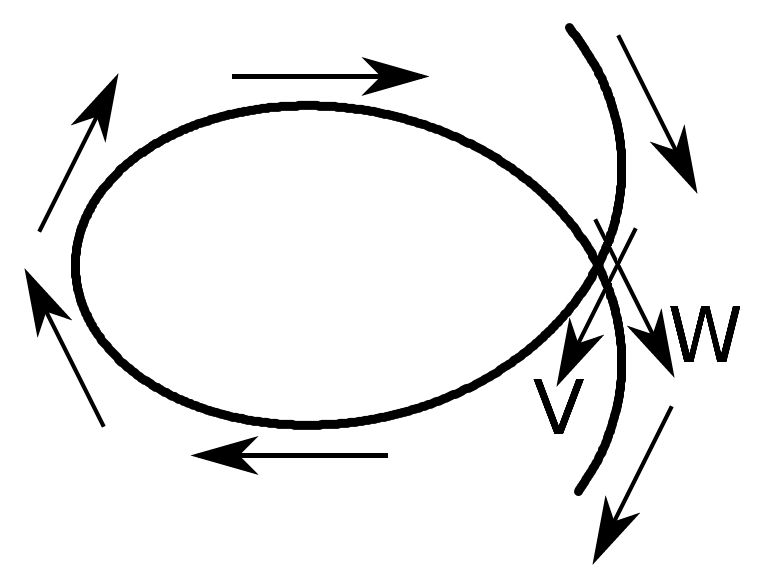}
    \caption{ Tangent vectors at a regular crossing-point of a curve.
      As we traverse the curve following the arrows from the
      top-right, we push-forward tangent vectors in the usual way.
      Thus, the first time the self-intersection is crossed, the
      tangent vector is $V$.  The second time, i.e. after traversing
      the loop in the clockwise direction, the tangent vector is
      $W$. Normal vectors are also well defined along the curve
      and uniquely specified once an outward direction is specified at
      any point.  In our specific example, we say that at the north
      pole, the outward direction is the $+z$ direction.  }
    \label{fig:knot}
\end{figure}

Before proceeding further, it might be useful to clarify the nature of
the MOTS with self-intersections shown in the bottom right panel of
Fig.~\ref{fig:merger1}.  Viewed as a submanifold of the 3-dimensional
Riemannian spatial slice $\Sigma$, this manifold might appear to be
non-differentiable at the point of self-intersection and one might be
concerned that there is no well defined normal to the manifold at that
point (and hence no well defined expansion either).  This is however
incorrect, and formally the curve is simply understood as an
\emph{immersion} instead of an embedding.  In the present case,
because of axisymmetry, we can restrict ourselves to a two-dimensional
section (say the $x$-$z$ plane as we have been using so far).  Then the
horizon is simply a parameterized curve, i.e. a mapping of the circle
$S^1$ into $\Sigma$, $f:S^1\rightarrow \Sigma$ (this is precisely how
this curve is defined numerically).  Using the map $f$, we can push
forward tangent vectors to $\Sigma$ and thus we have well defined
normals depending on which direction one traverses the point of
self-intersection (see Fig.~\ref{fig:knot}).  The relevant topological
property of the curve is the winding number, i.e. the number of
rotations that a tangent vector undergoes when we go all the way
around the curve; each loop adds +1 to the winding number.  Curves
with different winding numbers cannot be smoothly deformed into each
other \cite{CM_1937__4__276_0}.  This is why in order to get the
self-intersections, it is necessary to go through the cusp at the
merger.  In non-axisymmetric situations, we have to necessarily deal
with mappings of $S^2$ into the 3-manifold $\Sigma$ (which are in fact
simpler \cite{10.2307/1993205}), but we shall not discuss this here.

\subsection{Convergence}
\label{subsec:convergence}

Except for the modifications introduced earlier in
Sec.~\ref{sec:motsfinder}, we employ the same basic Newton-Kantorovich
search as in \cite{Pook-Kolb:2018igu} with each step being performed
using a pseudospectral method.  If the nonlinear search converges, we
expect the exponential convergence of the individual pseudospectral
steps to carry over to the solution of the full nonlinear problem.
\begin{figure}[h]
    \centering
    \includegraphics[width=0.48\textwidth]{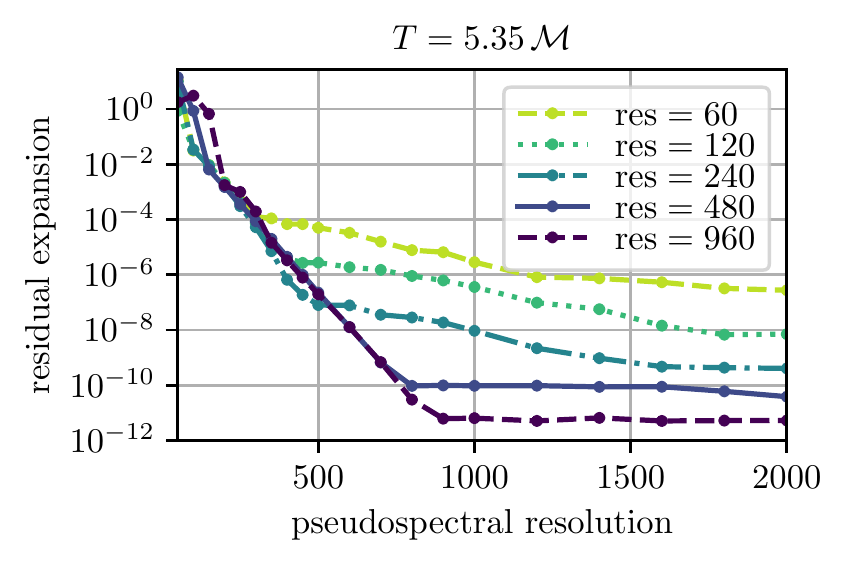}
    \includegraphics[width=0.48\textwidth]{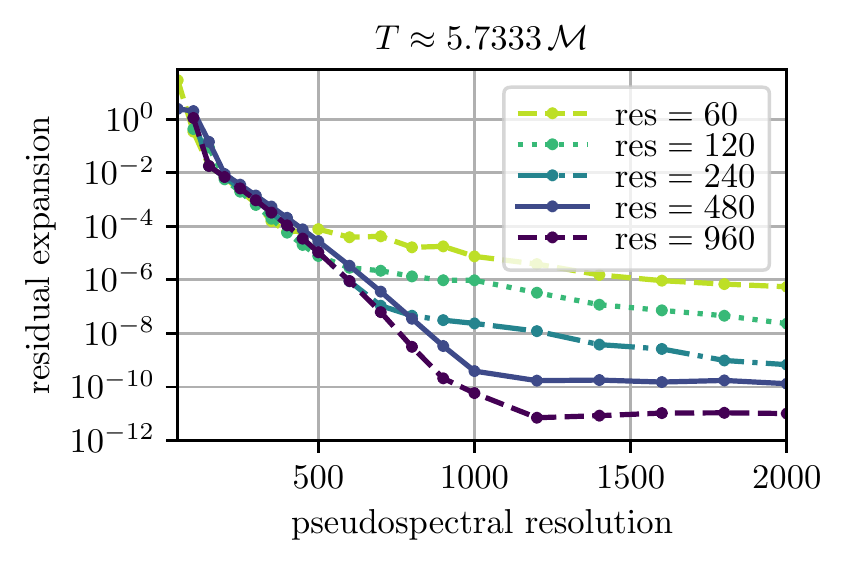}
    \caption{%
        Convergence of the residual expansion of $\Sin$ at one time
        before (upper panel) and after (lower panel) the individual
        horizons touch. Note that the inner common MOTS has
        self-intersections in the latter case.
        We plot the maximum absolute residual expansion between the
        collocation points over the pseudospectral resolution used to
        find the MOTS. This is independent of the grid resolution
        ${\rm res} = 1/h$ of the simulation.
        Exponential convergence is clearly visible up to reaching the
        plateau in the various cases.
        The plots also show that the plateau moves downward with
        increased grid resolution and that at lower resolution, we can
        identify a nonzero negative slope within the plateau,
        indicating the overfitting effect mentioned in the text.
    }
    \label{fig:spectral_convergence}
\end{figure}
This is indeed the case, as can be seen in
Fig.~\ref{fig:spectral_convergence}. It shows the maximum residual
expansion between the collocation points for $\Sin$ at two different
times of the simulation: one at $T=5.35\Munit$, where the MOTS is already
highly distorted, and one at $T\approx5.7333\Munit$. This second case is
{\em after} the individual MOTSs touch. At this stage, $\Sin$ lies in
the inside of $\Sone \cup \Stwo$ and intersects itself.
There is no qualitative difference in convergence and the plateau is
approximately at the same level for the same grid resolution.

We also see in Fig.~\ref{fig:spectral_convergence} that the
negative slope continues into the plateau region.  This effect is
more pronounced for lower grid resolutions and not noticeable for
$1/h=960$.  It is caused by fitting the horizon to features introduced
by the interpolation.  We avoid this unphysical effect in practice by
limiting the pseudospectral resolution as described at the end of
Sec.~\ref{subsec:interpolation}.

\begin{figure}[h]
    \centering
    \includegraphics[width=0.48\textwidth]{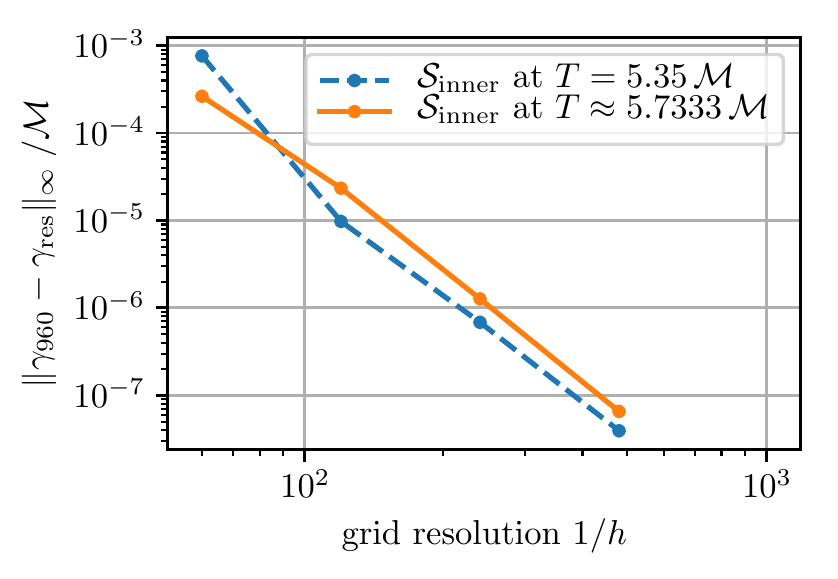}
    \caption{%
        Convergence of the coordinate shapes of the MOTSs for
        increasing numerical resolutions
        $1/h = 60$, $120$, $240$, $480$, $960$.
        Shown is the maximum coordinate distance of the horizons found
        in lower resolution simulations to the respective horizon
        found for $1/h = 960$.
    }
    \label{fig:curve_distances}
\end{figure}

Instead of varying the pseudospectral resolution, we can test
convergence for different grid resolutions $1/h$ of the simulation.
The quantity we use here is the convergence of the coordinate shapes
of the curves representing the MOTSs.  Fig.~\ref{fig:curve_distances}
shows that we indeed find convergence of the shapes.

\begin{figure*}[h]
    \centering
    \includegraphics[width=\textwidth]{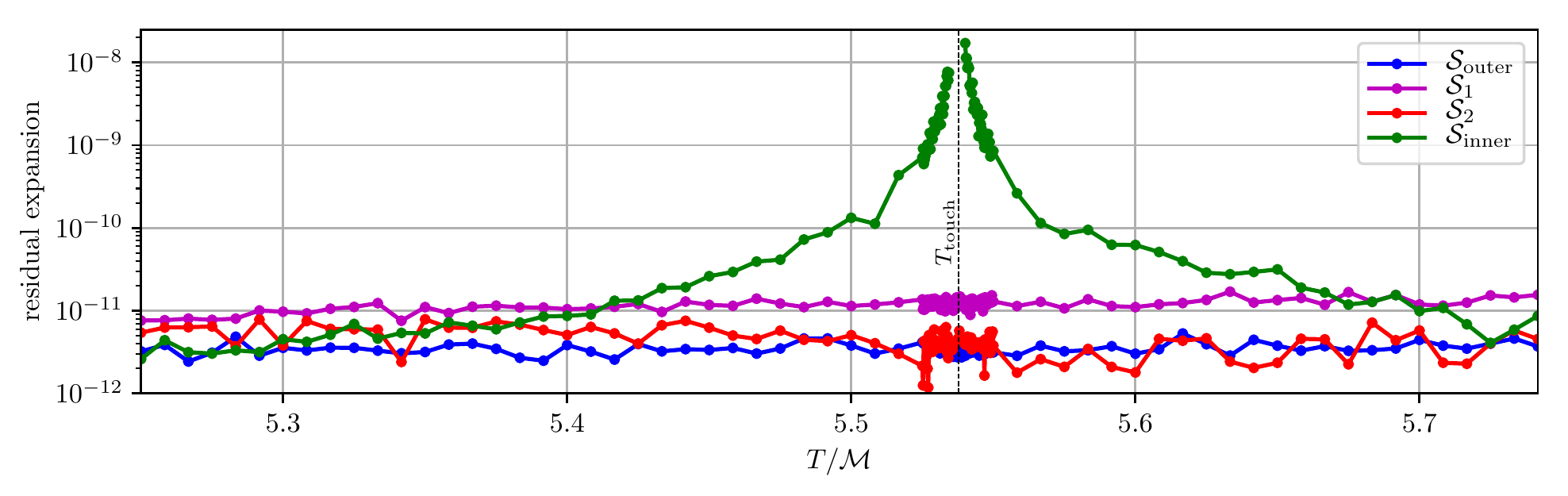}
    \caption{ The residual expansion of the various MOTSs
      ($\Surf_{1,2}$ and $\Surf_{\insubscript,\outsubscript}$) for the highest
      resolution $1/h=960$ as a function of time.  We plot the
      absolute maximum expansion sampled between the collocation
      points used by our pseudospectral method. }
    \label{fig:residual-expansion}
\end{figure*}

We show in Fig.~\ref{fig:residual-expansion}, as a function of time,
the residual expansion of the various MOTSs for the highest resolution
that we have considered, namely $1/h = 960$.  The residual expansion
is one of the key ingredients which gives us confidence that the
surfaces we find are indeed MOTSs.  Note first that for all the
``easy'' cases, namely for the two individual MOTSs
$\Surf_{1,2}$ and for the apparent horizon, the residual
expansion is no more than $\mathcal{O}(10^{-11})$. These horizons
do not have any portions with extreme curvatures and there is no
difficulty in locating them.  In fact, the residual expansion is
largest for the smaller horizon, and is $\mathcal{O}(10^{-12})$ for
the larger horizon and the apparent horizon.  The difficult case is of
course the inner common horizon, which required the
various technical improvements detailed earlier. The most difficult
cases are those which have the narrow neck and correspondingly highly
curved portions. There is a small duration of time near $\ttouch$ where
we are not able to locate $\Sin$.  At all the
other times shown in the plot, the residual expansion is no more than
$\mathcal{O}(10^{-9})$.  In fact, away from $\ttouch$, the residual
expansion is as good as for the other MOTSs.  In particular, this
is true after $T \sim 5.7\Munit$.  At these times $\Sin$ has
developed self-intersections.  Thus, our confidence in the
existence of self-intersecting MOTSs is the same as our confidence in the
existence of the other MOTSs, which of course are already well
established.  

\subsection{Area and stability} 
\label{subsec:area}

\begin{figure}[h]
    \centering
    \includegraphics[width=0.48\textwidth]{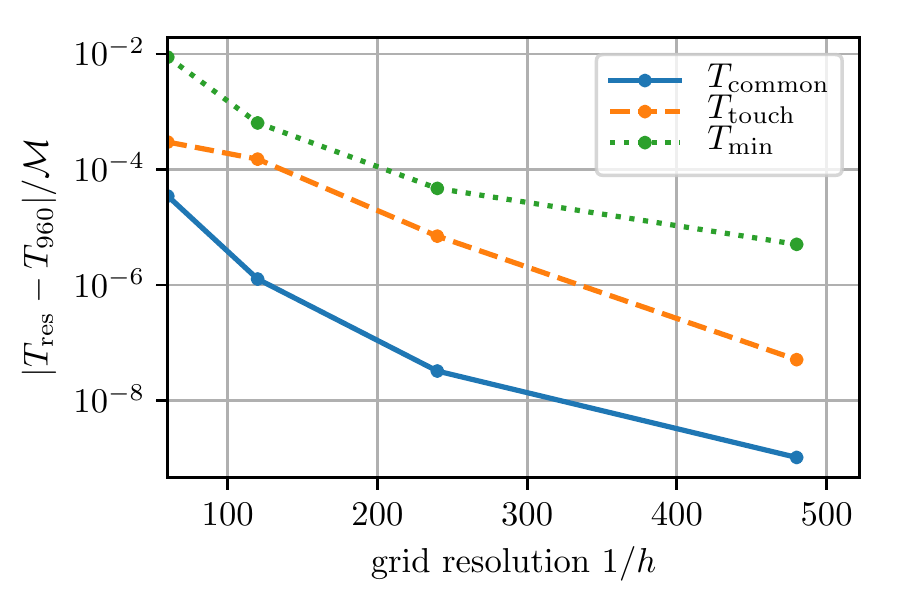}
    \caption{%
        Convergence of the various characteristic times with increased
        grid resolution. Shown is the difference between the value
        found at the finest resolution $1/h = 960$ and the respective
        lower resolution result. $\tAH$ is the time when the common
        horizon forms, $\Sone$ and $\Stwo$ touch at $\ttouch$, and the
        inner common horizon has a local minimal area at $\tmin$.
    }
    \label{fig:t_convergence}
\end{figure}

Some quantitative numbers for this evolution are:
\begin{itemize}
\item The common horizon forms at
  $\tAH \approx 1.37460222\Munit$.
\item The two individual horizons touch at
  $\ttouch \approx 5.5378176\Munit$.
\item The area of the inner horizon reaches a minimum at
  $\tmin \approx 5.50592\Munit$, i.e. just a little bit before
  $\ttouch$.  This behavior of $\Sin$ was previously noted in
  \cite{Pook-Kolb:2019iao}.
\end{itemize}
These values were computed at the various resolutions up to
$1/h = 960$ and converge up to the shown number of decimal places;
compare also Fig.~\ref{fig:t_convergence}.

\begin{figure*}[h]
    \centering
    \includegraphics[width=\textwidth]{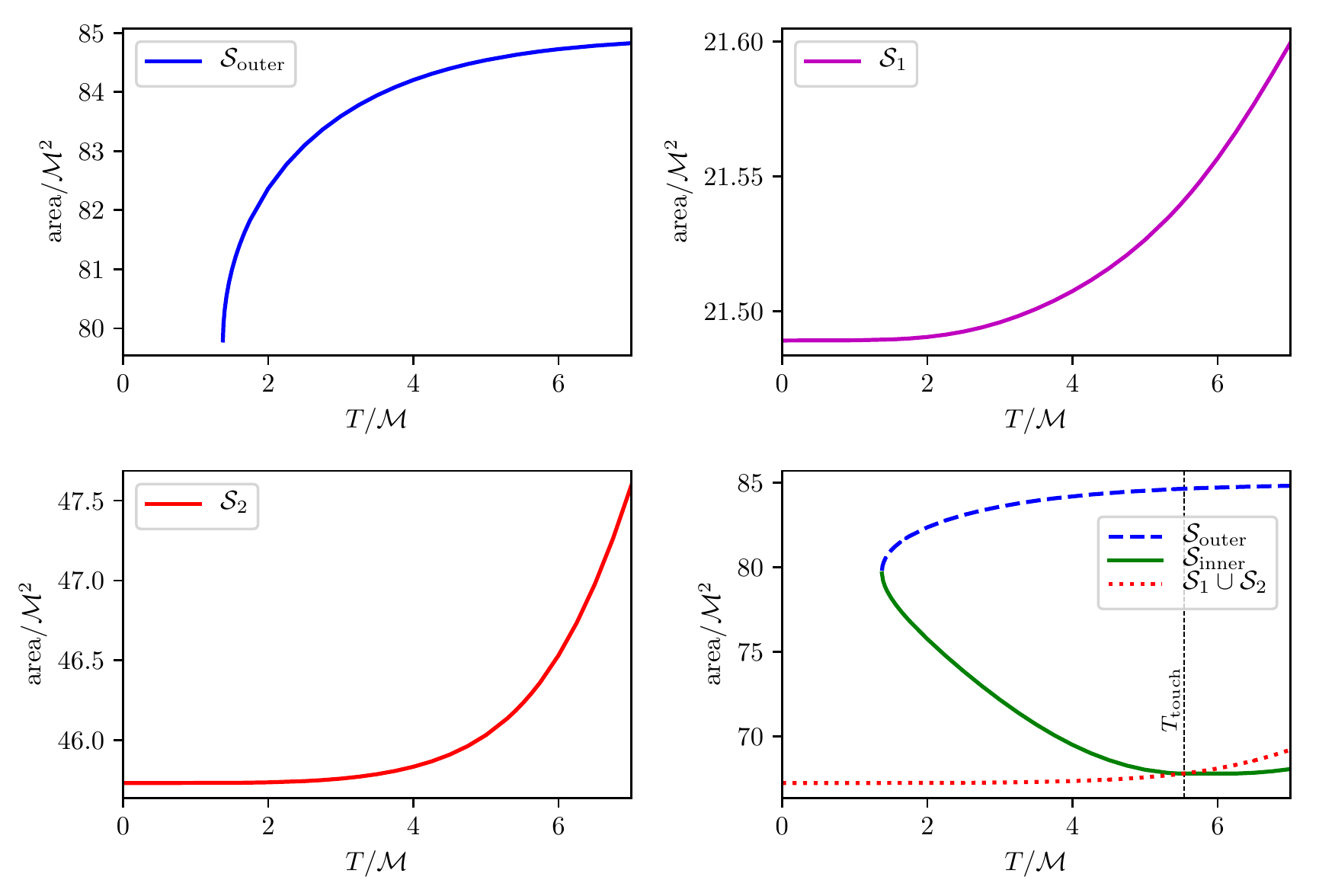}
    \caption{%
      Areas of the various horizons as functions of time.  The
      top-left panel shows the area of the apparent horizon which, as
      expected, asymptotes to a final constant value as the black hole
      reaches equilibrium. The top-right and bottom-left plots show
      the areas of the smaller and larger black holes respectively,
      both showing large increases at late times.  The area of the
      inner-common horizon is shown in the bottom-right panel. This
      panel also shows the apparent horizon, and the sum of the
      individual horizon areas.  }
    \label{fig:area-bl5}
\end{figure*}

The areas of the various horizons are plotted as functions of time in
Fig.~\ref{fig:area-bl5}.  The bottom-right panel presents a useful
picture of the merger process.  It shows the areas of the apparent
horizon, the inner common horizon and the sum of the areas of the
individual horizons.  It shows the formation and bifurcation of the
apparent horizon and it also shows the merger, i.e. the crossing of
the curves for the inner horizon and the individual horizons.

\begin{figure}[h]
    \centering
    \includegraphics[width=\columnwidth]{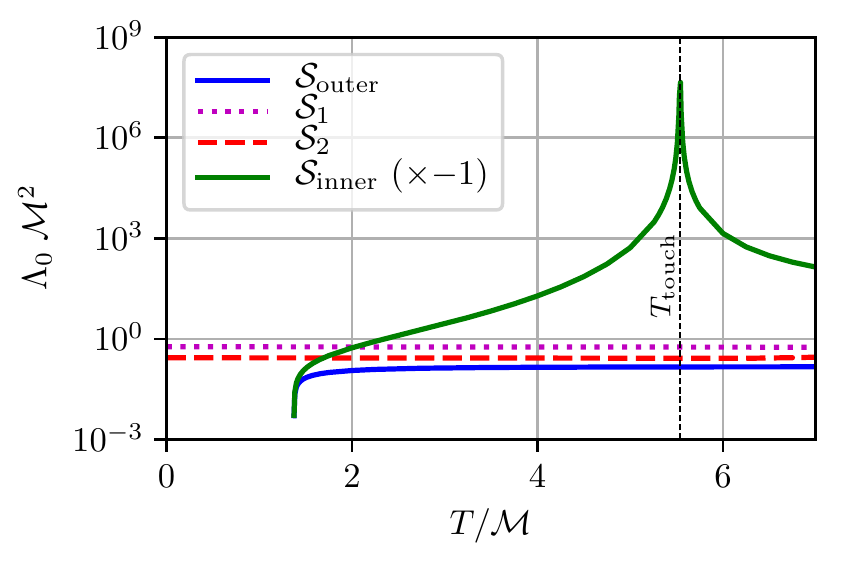}
    \caption{%
      First stability parameters $\Lambda_0$ (i.e. the principal
      eigenvalue of the stability operator) for the various horizons.
      $\Lambda_0$ is positive for all horizons except $\Sin$, for which
      we instead plot $-\Lambda_0$.  A cusp is clearly seen for the
      inner horizon. All the other horizons show unremarkable behavior
      in this respect; they remain stable as far as they can be
      reliably tracked.  }
    \label{fig:stability-bl5}
\end{figure}

\begin{figure}[h]
    \centering
    \includegraphics[width=\columnwidth]{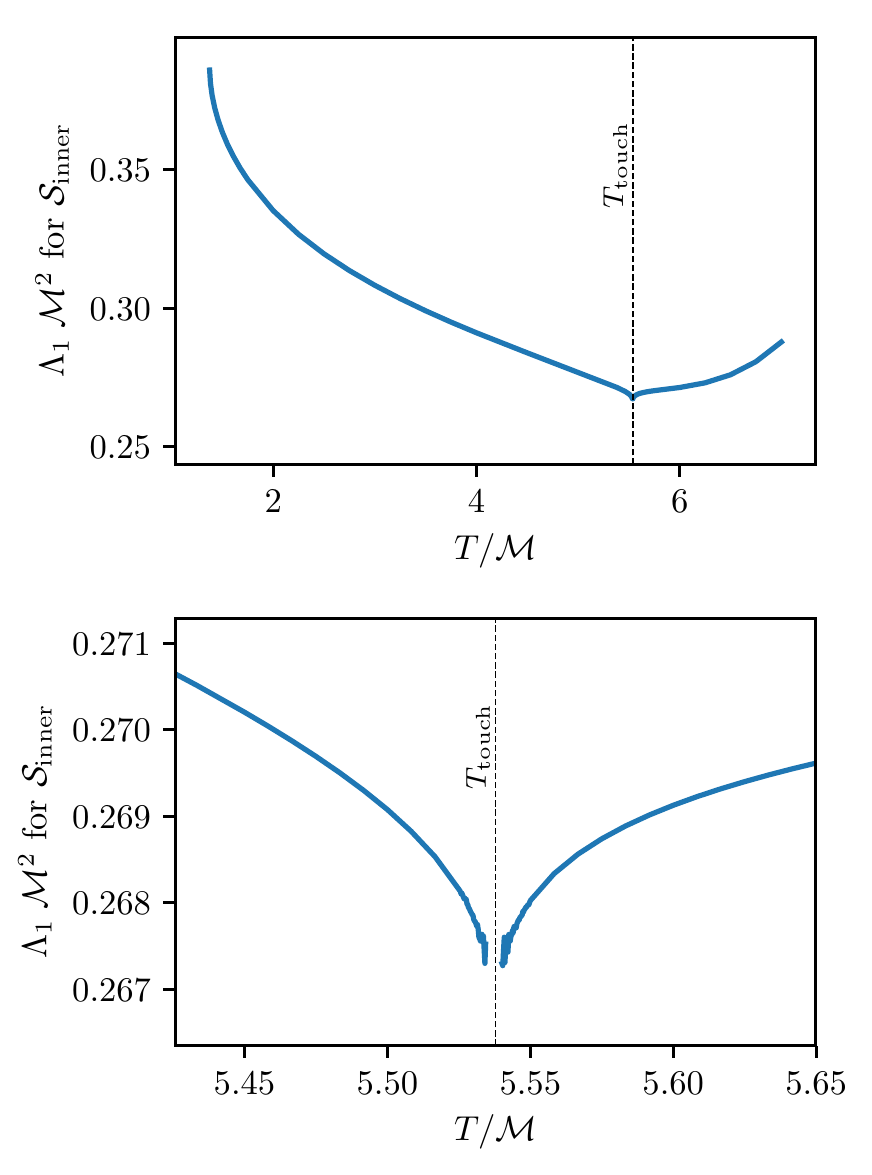}
    \caption{%
      The stability parameter for the inner horizon (upper panel) and
      zoom in around $\ttouch$ (lower panel).  We lose numerical
      precision very close to the merger time as the pseudospectral
      resolution becomes very large thereby increasing the condition
      number of the matrix of the discretized problem.
    }
    \label{fig:stability1-inner-bl5}
\end{figure}

The principal eigenvalue of the stability operator for the various
horizons is shown in Fig.~\ref{fig:stability-bl5}.  We see that
$\Lambda_0$ is always positive for $\Surf_{1,2}$ and for the apparent
horizon, and that it is not strongly varying.  $\Sin$ is more interesting.
When it is initially born, it coincides with the apparent horizon and
has $\Lambda_0=0$.  At all subsequent times, $\Sin$ has $\Lambda_0<0$;
to understand its stability we need to consider the next eigenvalue
$\Lambda_1$. But already from Fig.~\ref{fig:stability-bl5}, we see
interesting behavior of $\Lambda_0$ for $\Sin$, namely a cusp at
$\ttouch$.  Fig.~\ref{fig:stability1-inner-bl5} shows $\Lambda_1$ for
the inner horizon, and it is seen to be positive thus demonstrating
stability.  Again, we see a cusp-like behavior near $\ttouch$.

\section{Conclusions} 
\label{sec:conclusions}  

In this paper we examined in detail the scenario for the merger of
MOTSs outlined previously in \cite{Pook-Kolb:2019iao}.  We have done
this by evolving a particular Brill-Lindquist setup and finding all
MOTSs at various times.  We have tracked the inner common horizon with
high accuracy.  In particular, we present strong numerical evidence
that the inner horizon merges with the two individual horizons
precisely at the time when they touch.  Moreover, we find that the
inner horizon develops self-intersections just after the merger.  This
provides then a connected sequence of MOTSs taking us from the two
disjoint initial horizons to the final apparent horizon.  We have also
studied some basic properties of the MOTSs including their area and
stability.  There are numerous other interesting physical and
geometric properties of the world tube of MOTSs which shall be studied
in detail in forthcoming work.
 
\acknowledgments

We thank Abhay Ashtekar, Bernd Brugmann, Luis Lehner, and Andrey Shoom
for valuable discussions and comments.  We are especially grateful to
Jose-Luis Jaramillo for extensive discussions and for suggesting the
use of bipolar coordinates.

The MOTS finder
\cite{pook_kolb_daniel_2019_3260171}
used in this research is developed in Python with
\emph{SimulationIO}
\cite{erik_schnetter_2019_3258858}
being used for reading the numerical simulation data.  The libraries
\emph{SciPy}
\cite{Jones_SciPy},
\emph{NumPy}
\cite{van_der_Walt_NumPy},
\emph{mpmath}
\cite{mpmath},
\emph{SymPy}
\cite{meurer2017sympy}
and
\emph{Matplotlib}
\cite{Hunter:2007,michael_droettboom_2018_1202077}
were used for certain numerical, validation and plotting tasks.

O.B. acknowledges the National Science Foundation (NSF) for financial
support from Grant No. PHY-1607520.
This research was also supported by the Perimeter Institute for
Theoretical Physics.
Research at Perimeter Institute is supported in part by the Government
of Canada through the Department of Innovation, Science and Economic
Development Canada and by the Province of Ontario through the Ministry
of Economic Development, Job Creation and Trade.
%
This research was enabled in part by support provided by SciNet
(www.scinethpc.ca) and Compute Canada (www.computecanada.ca).
Computations were performed on the Niagara supercomputer at the SciNet
HPC Consortium \cite{Loken_2010}. SciNet is funded by: the Canada
Foundation for Innovation; the Government of Ontario; Ontario Research
Fund -- Research Excellence; and the University of Toronto.

\bibliography{mots_merger}{}

\begin{thebibliography}{74}%
\makeatletter
\providecommand \@ifxundefined [1]{%
 \@ifx{#1\undefined}
}%
\providecommand \@ifnum [1]{%
 \ifnum #1\expandafter \@firstoftwo
 \else \expandafter \@secondoftwo
 \fi
}%
\providecommand \@ifx [1]{%
 \ifx #1\expandafter \@firstoftwo
 \else \expandafter \@secondoftwo
 \fi
}%
\providecommand \natexlab [1]{#1}%
\providecommand \enquote  [1]{``#1''}%
\providecommand \bibnamefont  [1]{#1}%
\providecommand \bibfnamefont [1]{#1}%
\providecommand \citenamefont [1]{#1}%
\providecommand \href@noop [0]{\@secondoftwo}%
\providecommand \href [0]{\begingroup \@sanitize@url \@href}%
\providecommand \@href[1]{\@@startlink{#1}\@@href}%
\providecommand \@@href[1]{\endgroup#1\@@endlink}%
\providecommand \@sanitize@url [0]{\catcode `\\12\catcode `\$12\catcode
  `\&12\catcode `\#12\catcode `\^12\catcode `\_12\catcode `\%12\relax}%
\providecommand \@@startlink[1]{}%
\providecommand \@@endlink[0]{}%
\providecommand \url  [0]{\begingroup\@sanitize@url \@url }%
\providecommand \@url [1]{\endgroup\@href {#1}{\urlprefix }}%
\providecommand \urlprefix  [0]{URL }%
\providecommand \Eprint [0]{\href }%
\providecommand \doibase [0]{http://dx.doi.org/}%
\providecommand \selectlanguage [0]{\@gobble}%
\providecommand \bibinfo  [0]{\@secondoftwo}%
\providecommand \bibfield  [0]{\@secondoftwo}%
\providecommand \translation [1]{[#1]}%
\providecommand \BibitemOpen [0]{}%
\providecommand \bibitemStop [0]{}%
\providecommand \bibitemNoStop [0]{.\EOS\space}%
\providecommand \EOS [0]{\spacefactor3000\relax}%
\providecommand \BibitemShut  [1]{\csname bibitem#1\endcsname}%
\let\auto@bib@innerbib\@empty
\bibitem [{\citenamefont {Abbott}\ \emph
  {et~al.}(2016{\natexlab{a}})\citenamefont {Abbott} \emph
  {et~al.}}]{Abbott:2016blz}%
  \BibitemOpen
  \bibfield  {author} {\bibinfo {author} {\bibfnamefont {B.~P.}\ \bibnamefont
  {Abbott}} \emph {et~al.} (\bibinfo {collaboration} {LIGO Scientific,
  Virgo}),\ }\bibfield  {title} {\enquote {\bibinfo {title} {{Observation of
  Gravitational Waves from a Binary Black Hole Merger}},}\ }\href {\doibase
  10.1103/PhysRevLett.116.061102} {\bibfield  {journal} {\bibinfo  {journal}
  {Phys. Rev. Lett.}\ }\textbf {\bibinfo {volume} {116}},\ \bibinfo {pages}
  {061102} (\bibinfo {year} {2016}{\natexlab{a}})},\ \Eprint
  {http://arxiv.org/abs/1602.03837} {arXiv:1602.03837 [gr-qc]} \BibitemShut
  {NoStop}%
\bibitem [{\citenamefont {Abbott}\ \emph
  {et~al.}(2016{\natexlab{b}})\citenamefont {Abbott} \emph
  {et~al.}}]{TheLIGOScientific:2016pea}%
  \BibitemOpen
  \bibfield  {author} {\bibinfo {author} {\bibfnamefont {B.~P.}\ \bibnamefont
  {Abbott}} \emph {et~al.} (\bibinfo {collaboration} {LIGO Scientific,
  Virgo}),\ }\bibfield  {title} {\enquote {\bibinfo {title} {{Binary Black Hole
  Mergers in the first Advanced LIGO Observing Run}},}\ }\href {\doibase
  10.1103/PhysRevX.6.041015, 10.1103/PhysRevX.8.039903} {\bibfield  {journal}
  {\bibinfo  {journal} {Phys. Rev.}\ }\textbf {\bibinfo {volume} {X6}},\
  \bibinfo {pages} {041015} (\bibinfo {year} {2016}{\natexlab{b}})},\ \bibinfo
  {note} {[erratum: Phys. Rev.X8,no.3,039903(2018)]},\ \Eprint
  {http://arxiv.org/abs/1606.04856} {arXiv:1606.04856 [gr-qc]} \BibitemShut
  {NoStop}%
\bibitem [{\citenamefont {Abbott}\ \emph {et~al.}(2018)\citenamefont {Abbott}
  \emph {et~al.}}]{LIGOScientific:2018mvr}%
  \BibitemOpen
  \bibfield  {author} {\bibinfo {author} {\bibfnamefont {B.~P.}\ \bibnamefont
  {Abbott}} \emph {et~al.} (\bibinfo {collaboration} {LIGO Scientific,
  Virgo}),\ }\bibfield  {title} {\enquote {\bibinfo {title} {{GWTC-1: A
  Gravitational-Wave Transient Catalog of Compact Binary Mergers Observed by
  LIGO and Virgo during the First and Second Observing Runs}},}\ }\href@noop {}
  {\  (\bibinfo {year} {2018})},\ \Eprint {http://arxiv.org/abs/1811.12907}
  {arXiv:1811.12907 [astro-ph.HE]} \BibitemShut {NoStop}%
\bibitem [{\citenamefont {Green}\ and\ \citenamefont
  {Moffat}(2018)}]{Green:2017voq}%
  \BibitemOpen
  \bibfield  {author} {\bibinfo {author} {\bibfnamefont {Martin~A.}\
  \bibnamefont {Green}}\ and\ \bibinfo {author} {\bibfnamefont {J.~W.}\
  \bibnamefont {Moffat}},\ }\bibfield  {title} {\enquote {\bibinfo {title}
  {{Extraction of black hole coalescence waveforms from noisy data}},}\ }\href
  {\doibase 10.1016/j.physletb.2018.08.009} {\bibfield  {journal} {\bibinfo
  {journal} {Phys. Lett.}\ }\textbf {\bibinfo {volume} {B784}},\ \bibinfo
  {pages} {312--323} (\bibinfo {year} {2018})},\ \Eprint
  {http://arxiv.org/abs/1711.00347} {arXiv:1711.00347 [astro-ph.IM]}
  \BibitemShut {NoStop}%
\bibitem [{\citenamefont {Zackay}\ \emph {et~al.}(2019)\citenamefont {Zackay},
  \citenamefont {Venumadhav}, \citenamefont {Dai}, \citenamefont {Roulet},\
  and\ \citenamefont {Zaldarriaga}}]{Zackay:2019tzo}%
  \BibitemOpen
  \bibfield  {author} {\bibinfo {author} {\bibfnamefont {Barak}\ \bibnamefont
  {Zackay}}, \bibinfo {author} {\bibfnamefont {Tejaswi}\ \bibnamefont
  {Venumadhav}}, \bibinfo {author} {\bibfnamefont {Liang}\ \bibnamefont {Dai}},
  \bibinfo {author} {\bibfnamefont {Javier}\ \bibnamefont {Roulet}}, \ and\
  \bibinfo {author} {\bibfnamefont {Matias}\ \bibnamefont {Zaldarriaga}},\
  }\bibfield  {title} {\enquote {\bibinfo {title} {{A Highly Spinning and
  Aligned Binary Black Hole Merger in the Advanced LIGO First Observing
  Run}},}\ }\href@noop {} {\  (\bibinfo {year} {2019})},\ \Eprint
  {http://arxiv.org/abs/1902.10331} {arXiv:1902.10331 [astro-ph.HE]}
  \BibitemShut {NoStop}%
\bibitem [{\citenamefont {Nitz}\ \emph {et~al.}(2019)\citenamefont {Nitz},
  \citenamefont {Capano}, \citenamefont {Nielsen}, \citenamefont {Reyes},
  \citenamefont {White}, \citenamefont {Brown},\ and\ \citenamefont
  {Krishnan}}]{Nitz:2018imz}%
  \BibitemOpen
  \bibfield  {author} {\bibinfo {author} {\bibfnamefont {Alexander~H.}\
  \bibnamefont {Nitz}}, \bibinfo {author} {\bibfnamefont {Collin}\ \bibnamefont
  {Capano}}, \bibinfo {author} {\bibfnamefont {Alex~B.}\ \bibnamefont
  {Nielsen}}, \bibinfo {author} {\bibfnamefont {Steven}\ \bibnamefont {Reyes}},
  \bibinfo {author} {\bibfnamefont {Rebecca}\ \bibnamefont {White}}, \bibinfo
  {author} {\bibfnamefont {Duncan~A.}\ \bibnamefont {Brown}}, \ and\ \bibinfo
  {author} {\bibfnamefont {Badri}\ \bibnamefont {Krishnan}},\ }\bibfield
  {title} {\enquote {\bibinfo {title} {{1-OGC: The first open
  gravitational-wave catalog of binary mergers from analysis of public Advanced
  LIGO data}},}\ }\href {\doibase 10.3847/1538-4357/ab0108} {\bibfield
  {journal} {\bibinfo  {journal} {Astrophys. J.}\ }\textbf {\bibinfo {volume}
  {872}},\ \bibinfo {pages} {195} (\bibinfo {year} {2019})},\ \Eprint
  {http://arxiv.org/abs/1811.01921} {arXiv:1811.01921 [gr-qc]} \BibitemShut
  {NoStop}%
\bibitem [{\citenamefont {Venumadhav}\ \emph {et~al.}(2019)\citenamefont
  {Venumadhav}, \citenamefont {Zackay}, \citenamefont {Roulet}, \citenamefont
  {Dai},\ and\ \citenamefont {Zaldarriaga}}]{Venumadhav:2019lyq}%
  \BibitemOpen
  \bibfield  {author} {\bibinfo {author} {\bibfnamefont {Tejaswi}\ \bibnamefont
  {Venumadhav}}, \bibinfo {author} {\bibfnamefont {Barak}\ \bibnamefont
  {Zackay}}, \bibinfo {author} {\bibfnamefont {Javier}\ \bibnamefont {Roulet}},
  \bibinfo {author} {\bibfnamefont {Liang}\ \bibnamefont {Dai}}, \ and\
  \bibinfo {author} {\bibfnamefont {Matias}\ \bibnamefont {Zaldarriaga}},\
  }\bibfield  {title} {\enquote {\bibinfo {title} {{New Binary Black Hole
  Mergers in the Second Observing Run of Advanced LIGO and Advanced Virgo}},}\
  }\href@noop {} {\  (\bibinfo {year} {2019})},\ \Eprint
  {http://arxiv.org/abs/1904.07214} {arXiv:1904.07214 [astro-ph.HE]}
  \BibitemShut {NoStop}%
\bibitem [{\citenamefont {Gupta}\ \emph {et~al.}(2018)\citenamefont {Gupta},
  \citenamefont {Krishnan}, \citenamefont {Nielsen},\ and\ \citenamefont
  {Schnetter}}]{Gupta:2018znn}%
  \BibitemOpen
  \bibfield  {author} {\bibinfo {author} {\bibfnamefont {Anshu}\ \bibnamefont
  {Gupta}}, \bibinfo {author} {\bibfnamefont {Badri}\ \bibnamefont {Krishnan}},
  \bibinfo {author} {\bibfnamefont {Alex}\ \bibnamefont {Nielsen}}, \ and\
  \bibinfo {author} {\bibfnamefont {Erik}\ \bibnamefont {Schnetter}},\
  }\bibfield  {title} {\enquote {\bibinfo {title} {{Dynamics of marginally
  trapped surfaces in a binary black hole merger: Growth and approach to
  equilibrium}},}\ }\href {\doibase 10.1103/PhysRevD.97.084028} {\bibfield
  {journal} {\bibinfo  {journal} {Phys. Rev.}\ }\textbf {\bibinfo {volume}
  {D97}},\ \bibinfo {pages} {084028} (\bibinfo {year} {2018})},\ \Eprint
  {http://arxiv.org/abs/1801.07048} {arXiv:1801.07048 [gr-qc]} \BibitemShut
  {NoStop}%
\bibitem [{\citenamefont {Jaramillo}\ \emph {et~al.}(2012)\citenamefont
  {Jaramillo}, \citenamefont {Macedo}, \citenamefont {M{\"o}sta},\ and\
  \citenamefont {Rezzolla}}]{Jaramillo:2011rf}%
  \BibitemOpen
  \bibfield  {author} {\bibinfo {author} {\bibfnamefont {Jose~Luis}\
  \bibnamefont {Jaramillo}}, \bibinfo {author} {\bibfnamefont {Rodrigo~P.}\
  \bibnamefont {Macedo}}, \bibinfo {author} {\bibfnamefont {Philipp}\
  \bibnamefont {M{\"o}sta}}, \ and\ \bibinfo {author} {\bibfnamefont {Luciano}\
  \bibnamefont {Rezzolla}},\ }\bibfield  {title} {\enquote {\bibinfo {title}
  {{Black-hole horizons as probes of black-hole dynamics II: geometrical
  insights}},}\ }\href {\doibase 10.1103/PhysRevD.85.084031} {\bibfield
  {journal} {\bibinfo  {journal} {Phys. Rev.}\ }\textbf {\bibinfo {volume}
  {D85}},\ \bibinfo {pages} {084031} (\bibinfo {year} {2012})},\ \Eprint
  {http://arxiv.org/abs/1108.0061} {arXiv:1108.0061 [gr-qc]} \BibitemShut
  {NoStop}%
\bibitem [{\citenamefont {Jaramillo}\ \emph {et~al.}(2011)\citenamefont
  {Jaramillo}, \citenamefont {Macedo}, \citenamefont {M{\"o}sta},\ and\
  \citenamefont {Rezzolla}}]{Jaramillo:2012rr}%
  \BibitemOpen
  \bibfield  {author} {\bibinfo {author} {\bibfnamefont {J.~L.}\ \bibnamefont
  {Jaramillo}}, \bibinfo {author} {\bibfnamefont {R.~P.}\ \bibnamefont
  {Macedo}}, \bibinfo {author} {\bibfnamefont {P.}~\bibnamefont {M{\"o}sta}}, \
  and\ \bibinfo {author} {\bibfnamefont {L.}~\bibnamefont {Rezzolla}},\
  }\bibfield  {title} {\enquote {\bibinfo {title} {{Towards a cross-correlation
  approach to strong-field dynamics in Black Hole spacetimes}},}\ }\bibfield
  {booktitle} {\emph {\bibinfo {booktitle} {{Proceedings, Spanish Relativity
  Meeting : Towards new paradigms. (ERE 2011): Madrid, Spain, August
  29-September 2, 2011}}},\ }\href {\doibase 10.1063/1.4734411} {\bibfield
  {journal} {\bibinfo  {journal} {AIP Conf. Proc.}\ }\textbf {\bibinfo {volume}
  {1458}},\ \bibinfo {pages} {158--173} (\bibinfo {year} {2011})},\ \Eprint
  {http://arxiv.org/abs/1205.3902} {arXiv:1205.3902 [gr-qc]} \BibitemShut
  {NoStop}%
\bibitem [{\citenamefont {Kamaretsos}\ \emph
  {et~al.}(2012{\natexlab{a}})\citenamefont {Kamaretsos}, \citenamefont
  {Hannam},\ and\ \citenamefont {Sathyaprakash}}]{Kamaretsos:2012bs}%
  \BibitemOpen
  \bibfield  {author} {\bibinfo {author} {\bibfnamefont {Ioannis}\ \bibnamefont
  {Kamaretsos}}, \bibinfo {author} {\bibfnamefont {Mark}\ \bibnamefont
  {Hannam}}, \ and\ \bibinfo {author} {\bibfnamefont {B.}~\bibnamefont
  {Sathyaprakash}},\ }\bibfield  {title} {\enquote {\bibinfo {title} {{Is
  black-hole ringdown a memory of its progenitor?}}}\ }\href {\doibase
  10.1103/PhysRevLett.109.141102} {\bibfield  {journal} {\bibinfo  {journal}
  {Phys. Rev. Lett.}\ }\textbf {\bibinfo {volume} {109}},\ \bibinfo {pages}
  {141102} (\bibinfo {year} {2012}{\natexlab{a}})},\ \Eprint
  {http://arxiv.org/abs/1207.0399} {arXiv:1207.0399 [gr-qc]} \BibitemShut
  {NoStop}%
\bibitem [{\citenamefont {Kamaretsos}\ \emph
  {et~al.}(2012{\natexlab{b}})\citenamefont {Kamaretsos}, \citenamefont
  {Hannam}, \citenamefont {Husa},\ and\ \citenamefont
  {Sathyaprakash}}]{Kamaretsos:2011um}%
  \BibitemOpen
  \bibfield  {author} {\bibinfo {author} {\bibfnamefont {Ioannis}\ \bibnamefont
  {Kamaretsos}}, \bibinfo {author} {\bibfnamefont {Mark}\ \bibnamefont
  {Hannam}}, \bibinfo {author} {\bibfnamefont {Sascha}\ \bibnamefont {Husa}}, \
  and\ \bibinfo {author} {\bibfnamefont {B.~S.}\ \bibnamefont
  {Sathyaprakash}},\ }\bibfield  {title} {\enquote {\bibinfo {title}
  {{Black-hole hair loss: learning about binary progenitors from ringdown
  signals}},}\ }\href {\doibase 10.1103/PhysRevD.85.024018} {\bibfield
  {journal} {\bibinfo  {journal} {Phys. Rev.}\ }\textbf {\bibinfo {volume}
  {D85}},\ \bibinfo {pages} {024018} (\bibinfo {year} {2012}{\natexlab{b}})},\
  \Eprint {http://arxiv.org/abs/1107.0854} {arXiv:1107.0854 [gr-qc]}
  \BibitemShut {NoStop}%
\bibitem [{\citenamefont {Bhagwat}\ \emph {et~al.}(2018)\citenamefont
  {Bhagwat}, \citenamefont {Okounkova}, \citenamefont {Ballmer}, \citenamefont
  {Brown}, \citenamefont {Giesler}, \citenamefont {Scheel},\ and\ \citenamefont
  {Teukolsky}}]{Bhagwat:2017tkm}%
  \BibitemOpen
  \bibfield  {author} {\bibinfo {author} {\bibfnamefont {Swetha}\ \bibnamefont
  {Bhagwat}}, \bibinfo {author} {\bibfnamefont {Maria}\ \bibnamefont
  {Okounkova}}, \bibinfo {author} {\bibfnamefont {Stefan~W.}\ \bibnamefont
  {Ballmer}}, \bibinfo {author} {\bibfnamefont {Duncan~A.}\ \bibnamefont
  {Brown}}, \bibinfo {author} {\bibfnamefont {Matthew}\ \bibnamefont
  {Giesler}}, \bibinfo {author} {\bibfnamefont {Mark~A.}\ \bibnamefont
  {Scheel}}, \ and\ \bibinfo {author} {\bibfnamefont {Saul~A.}\ \bibnamefont
  {Teukolsky}},\ }\bibfield  {title} {\enquote {\bibinfo {title} {{On choosing
  the start time of binary black hole ringdowns}},}\ }\href {\doibase
  10.1103/PhysRevD.97.104065} {\bibfield  {journal} {\bibinfo  {journal} {Phys.
  Rev.}\ }\textbf {\bibinfo {volume} {D97}},\ \bibinfo {pages} {104065}
  (\bibinfo {year} {2018})},\ \Eprint {http://arxiv.org/abs/1711.00926}
  {arXiv:1711.00926 [gr-qc]} \BibitemShut {NoStop}%
\bibitem [{\citenamefont {Borhanian}\ \emph {et~al.}(2019)\citenamefont
  {Borhanian}, \citenamefont {Arun}, \citenamefont {Pfeiffer},\ and\
  \citenamefont {Sathyaprakash}}]{Borhanian:2019kxt}%
  \BibitemOpen
  \bibfield  {author} {\bibinfo {author} {\bibfnamefont {Ssohrab}\ \bibnamefont
  {Borhanian}}, \bibinfo {author} {\bibfnamefont {K.~G.}\ \bibnamefont {Arun}},
  \bibinfo {author} {\bibfnamefont {Harald~P.}\ \bibnamefont {Pfeiffer}}, \
  and\ \bibinfo {author} {\bibfnamefont {B.~S.}\ \bibnamefont
  {Sathyaprakash}},\ }\bibfield  {title} {\enquote {\bibinfo {title}
  {{Signature of horizon dynamics in binary black hole gravitational
  waveforms}},}\ }\href@noop {} {\  (\bibinfo {year} {2019})},\ \Eprint
  {http://arxiv.org/abs/1901.08516} {arXiv:1901.08516 [gr-qc]} \BibitemShut
  {NoStop}%
\bibitem [{\citenamefont {Pretorius}(2005{\natexlab{a}})}]{Pretorius:2005gq}%
  \BibitemOpen
  \bibfield  {author} {\bibinfo {author} {\bibfnamefont {Frans}\ \bibnamefont
  {Pretorius}},\ }\bibfield  {title} {\enquote {\bibinfo {title} {{Evolution of
  Binary Black Hole Spacetimes}},}\ }\href {\doibase
  10.1103/PhysRevLett.95.121101} {\bibfield  {journal} {\bibinfo  {journal}
  {Phys. Rev. Lett.}\ }\textbf {\bibinfo {volume} {95}},\ \bibinfo {pages}
  {121101} (\bibinfo {year} {2005}{\natexlab{a}})},\ \Eprint
  {http://arxiv.org/abs/gr-qc/0507014} {arXiv:gr-qc/0507014} \BibitemShut
  {NoStop}%
\bibitem [{\citenamefont {Campanelli}\ \emph {et~al.}(2006)\citenamefont
  {Campanelli}, \citenamefont {Lousto}, \citenamefont {Marronetti},\ and\
  \citenamefont {Zlochower}}]{Campanelli:2005dd}%
  \BibitemOpen
  \bibfield  {author} {\bibinfo {author} {\bibfnamefont {Manuela}\ \bibnamefont
  {Campanelli}}, \bibinfo {author} {\bibfnamefont {C.~O.}\ \bibnamefont
  {Lousto}}, \bibinfo {author} {\bibfnamefont {P.}~\bibnamefont {Marronetti}},
  \ and\ \bibinfo {author} {\bibfnamefont {Y.}~\bibnamefont {Zlochower}},\
  }\bibfield  {title} {\enquote {\bibinfo {title} {{Accurate evolutions of
  orbiting black-hole binaries without excision}},}\ }\href {\doibase
  10.1103/PhysRevLett.96.111101} {\bibfield  {journal} {\bibinfo  {journal}
  {Phys. Rev. Lett.}\ }\textbf {\bibinfo {volume} {96}},\ \bibinfo {pages}
  {111101} (\bibinfo {year} {2006})},\ \Eprint
  {http://arxiv.org/abs/gr-qc/0511048} {arXiv:gr-qc/0511048 [gr-qc]}
  \BibitemShut {NoStop}%
\bibitem [{\citenamefont {Baker}\ \emph {et~al.}(2006)\citenamefont {Baker},
  \citenamefont {Centrella}, \citenamefont {Choi}, \citenamefont {Koppitz},\
  and\ \citenamefont {van Meter}}]{Baker:2005vv}%
  \BibitemOpen
  \bibfield  {author} {\bibinfo {author} {\bibfnamefont {John~G.}\ \bibnamefont
  {Baker}}, \bibinfo {author} {\bibfnamefont {Joan}\ \bibnamefont {Centrella}},
  \bibinfo {author} {\bibfnamefont {Dae-Il}\ \bibnamefont {Choi}}, \bibinfo
  {author} {\bibfnamefont {Michael}\ \bibnamefont {Koppitz}}, \ and\ \bibinfo
  {author} {\bibfnamefont {James}\ \bibnamefont {van Meter}},\ }\bibfield
  {title} {\enquote {\bibinfo {title} {{Gravitational wave extraction from an
  inspiraling configuration of merging black holes}},}\ }\href {\doibase
  10.1103/PhysRevLett.96.111102} {\bibfield  {journal} {\bibinfo  {journal}
  {Phys. Rev. Lett.}\ }\textbf {\bibinfo {volume} {96}},\ \bibinfo {pages}
  {111102} (\bibinfo {year} {2006})},\ \Eprint
  {http://arxiv.org/abs/gr-qc/0511103} {arXiv:gr-qc/0511103 [gr-qc]}
  \BibitemShut {NoStop}%
\bibitem [{\citenamefont {Szilagyi}\ \emph {et~al.}(2009)\citenamefont
  {Szilagyi}, \citenamefont {Lindblom},\ and\ \citenamefont
  {Scheel}}]{Szilagyi:2009qz}%
  \BibitemOpen
  \bibfield  {author} {\bibinfo {author} {\bibfnamefont {Bela}\ \bibnamefont
  {Szilagyi}}, \bibinfo {author} {\bibfnamefont {Lee}\ \bibnamefont
  {Lindblom}}, \ and\ \bibinfo {author} {\bibfnamefont {Mark~A.}\ \bibnamefont
  {Scheel}},\ }\bibfield  {title} {\enquote {\bibinfo {title} {{Simulations of
  Binary Black Hole Mergers Using Spectral Methods}},}\ }\href {\doibase
  10.1103/PhysRevD.80.124010} {\bibfield  {journal} {\bibinfo  {journal} {Phys.
  Rev.}\ }\textbf {\bibinfo {volume} {D80}},\ \bibinfo {pages} {124010}
  (\bibinfo {year} {2009})},\ \Eprint {http://arxiv.org/abs/0909.3557}
  {arXiv:0909.3557 [gr-qc]} \BibitemShut {NoStop}%
\bibitem [{\citenamefont {Hughes}\ \emph {et~al.}(1994)\citenamefont {Hughes},
  \citenamefont {Keeton}, \citenamefont {Walker}, \citenamefont {Walsh},
  \citenamefont {Shapiro},\ and\ \citenamefont {Teukolsky}}]{Hughes:1994ea}%
  \BibitemOpen
  \bibfield  {author} {\bibinfo {author} {\bibfnamefont {Scott~A.}\
  \bibnamefont {Hughes}}, \bibinfo {author} {\bibfnamefont {Charles~R.}\
  \bibnamefont {Keeton}}, \bibinfo {author} {\bibfnamefont {Paul}\ \bibnamefont
  {Walker}}, \bibinfo {author} {\bibfnamefont {Kevin~T.}\ \bibnamefont
  {Walsh}}, \bibinfo {author} {\bibfnamefont {Stuart~L.}\ \bibnamefont
  {Shapiro}}, \ and\ \bibinfo {author} {\bibfnamefont {Saul~A.}\ \bibnamefont
  {Teukolsky}},\ }\bibfield  {title} {\enquote {\bibinfo {title} {{Finding
  black holes in numerical space-times}},}\ }\href {\doibase
  10.1103/PhysRevD.49.4004} {\bibfield  {journal} {\bibinfo  {journal} {Phys.
  Rev.}\ }\textbf {\bibinfo {volume} {D49}},\ \bibinfo {pages} {4004--4015}
  (\bibinfo {year} {1994})}\BibitemShut {NoStop}%
\bibitem [{\citenamefont {Diener}(2003)}]{Diener:2003jc}%
  \BibitemOpen
  \bibfield  {author} {\bibinfo {author} {\bibfnamefont {Peter}\ \bibnamefont
  {Diener}},\ }\bibfield  {title} {\enquote {\bibinfo {title} {{A New general
  purpose event horizon finder for 3-D numerical space-times}},}\ }\href
  {\doibase 10.1088/0264-9381/20/22/014} {\bibfield  {journal} {\bibinfo
  {journal} {Class.Quant.Grav.}\ }\textbf {\bibinfo {volume} {20}},\ \bibinfo
  {pages} {4901--4918} (\bibinfo {year} {2003})},\ \Eprint
  {http://arxiv.org/abs/gr-qc/0305039} {arXiv:gr-qc/0305039 [gr-qc]}
  \BibitemShut {NoStop}%
\bibitem [{\citenamefont {Anninos}\ \emph {et~al.}(1995)\citenamefont
  {Anninos}, \citenamefont {Bernstein}, \citenamefont {Brandt}, \citenamefont
  {Libson}, \citenamefont {Mass\'o}, \citenamefont {Seidel}, \citenamefont
  {Smarr}, \citenamefont {Suen},\ and\ \citenamefont
  {Walker}}]{PhysRevLett.74.630}%
  \BibitemOpen
  \bibfield  {author} {\bibinfo {author} {\bibfnamefont {Peter}\ \bibnamefont
  {Anninos}}, \bibinfo {author} {\bibfnamefont {David}\ \bibnamefont
  {Bernstein}}, \bibinfo {author} {\bibfnamefont {Steven}\ \bibnamefont
  {Brandt}}, \bibinfo {author} {\bibfnamefont {Joseph}\ \bibnamefont {Libson}},
  \bibinfo {author} {\bibfnamefont {Joan}\ \bibnamefont {Mass\'o}}, \bibinfo
  {author} {\bibfnamefont {Edward}\ \bibnamefont {Seidel}}, \bibinfo {author}
  {\bibfnamefont {Larry}\ \bibnamefont {Smarr}}, \bibinfo {author}
  {\bibfnamefont {Wai-Mo}\ \bibnamefont {Suen}}, \ and\ \bibinfo {author}
  {\bibfnamefont {Paul}\ \bibnamefont {Walker}},\ }\bibfield  {title} {\enquote
  {\bibinfo {title} {Dynamics of apparent and event horizons},}\ }\href
  {\doibase 10.1103/PhysRevLett.74.630} {\bibfield  {journal} {\bibinfo
  {journal} {Phys. Rev. Lett.}\ }\textbf {\bibinfo {volume} {74}},\ \bibinfo
  {pages} {630--633} (\bibinfo {year} {1995})}\BibitemShut {NoStop}%
\bibitem [{\citenamefont {Thornburg}(2007)}]{Thornburg:2006zb}%
  \BibitemOpen
  \bibfield  {author} {\bibinfo {author} {\bibfnamefont {Jonathan}\
  \bibnamefont {Thornburg}},\ }\bibfield  {title} {\enquote {\bibinfo {title}
  {{Event and Apparent Horizon Finders for $3+1$ Numerical Relativity}},}\
  }\href@noop {} {\bibfield  {journal} {\bibinfo  {journal} {Living Rev. Rel.}\
  }\textbf {\bibinfo {volume} {10}},\ \bibinfo {pages} {3} (\bibinfo {year}
  {2007})},\ \Eprint {http://arxiv.org/abs/gr-qc/0512169} {arXiv:gr-qc/0512169}
  \BibitemShut {NoStop}%
\bibitem [{\citenamefont {Matzner}\ \emph {et~al.}(1995)\citenamefont
  {Matzner}, \citenamefont {Seidel}, \citenamefont {Shapiro}, \citenamefont
  {Smarr}, \citenamefont {Suen}, \citenamefont {Teukolsky},\ and\ \citenamefont
  {Winicour}}]{Matzner:1995ib}%
  \BibitemOpen
  \bibfield  {author} {\bibinfo {author} {\bibfnamefont {R.~A.}\ \bibnamefont
  {Matzner}}, \bibinfo {author} {\bibfnamefont {H.~E.}\ \bibnamefont {Seidel}},
  \bibinfo {author} {\bibfnamefont {Stuart~L.}\ \bibnamefont {Shapiro}},
  \bibinfo {author} {\bibfnamefont {L.}~\bibnamefont {Smarr}}, \bibinfo
  {author} {\bibfnamefont {W.~M.}\ \bibnamefont {Suen}}, \bibinfo {author}
  {\bibfnamefont {Saul~A.}\ \bibnamefont {Teukolsky}}, \ and\ \bibinfo {author}
  {\bibfnamefont {J.}~\bibnamefont {Winicour}},\ }\bibfield  {title} {\enquote
  {\bibinfo {title} {{Geometry of a black hole collision}},}\ }\href {\doibase
  10.1126/science.270.5238.941} {\bibfield  {journal} {\bibinfo  {journal}
  {Science}\ }\textbf {\bibinfo {volume} {270}},\ \bibinfo {pages} {941--947}
  (\bibinfo {year} {1995})}\BibitemShut {NoStop}%
\bibitem [{\citenamefont {Husa}\ and\ \citenamefont
  {Winicour}(1999)}]{PhysRevD.60.084019}%
  \BibitemOpen
  \bibfield  {author} {\bibinfo {author} {\bibfnamefont {Sascha}\ \bibnamefont
  {Husa}}\ and\ \bibinfo {author} {\bibfnamefont {Jeffrey}\ \bibnamefont
  {Winicour}},\ }\bibfield  {title} {\enquote {\bibinfo {title} {Asymmetric
  merger of black holes},}\ }\href {\doibase 10.1103/PhysRevD.60.084019}
  {\bibfield  {journal} {\bibinfo  {journal} {Phys. Rev. D}\ }\textbf {\bibinfo
  {volume} {60}},\ \bibinfo {pages} {084019} (\bibinfo {year}
  {1999})}\BibitemShut {NoStop}%
\bibitem [{\citenamefont {Chrusciel}\ \emph {et~al.}(2002)\citenamefont
  {Chrusciel}, \citenamefont {Fu}, \citenamefont {Galloway},\ and\
  \citenamefont {Howard}}]{Chrusciel:2000gj}%
  \BibitemOpen
  \bibfield  {author} {\bibinfo {author} {\bibfnamefont {Piotr~T.}\
  \bibnamefont {Chrusciel}}, \bibinfo {author} {\bibfnamefont {Joseph H.~G.}\
  \bibnamefont {Fu}}, \bibinfo {author} {\bibfnamefont {Gregory~J.}\
  \bibnamefont {Galloway}}, \ and\ \bibinfo {author} {\bibfnamefont {Ralph}\
  \bibnamefont {Howard}},\ }\bibfield  {title} {\enquote {\bibinfo {title} {{On
  fine differentiability properties of horizons and applications to Riemannian
  geometry}},}\ }\href {\doibase 10.1016/S0393-0440(01)00044-4} {\bibfield
  {journal} {\bibinfo  {journal} {J. Geom. Phys.}\ }\textbf {\bibinfo {volume}
  {41}},\ \bibinfo {pages} {1--12} (\bibinfo {year} {2002})},\ \Eprint
  {http://arxiv.org/abs/gr-qc/0011067} {arXiv:gr-qc/0011067 [gr-qc]}
  \BibitemShut {NoStop}%
\bibitem [{\citenamefont {Chrusciel}\ and\ \citenamefont
  {Galloway}(1998)}]{Chrusciel:1996tw}%
  \BibitemOpen
  \bibfield  {author} {\bibinfo {author} {\bibfnamefont {Piotr~T.}\
  \bibnamefont {Chrusciel}}\ and\ \bibinfo {author} {\bibfnamefont
  {Gregory~J.}\ \bibnamefont {Galloway}},\ }\bibfield  {title} {\enquote
  {\bibinfo {title} {{`Nowhere' differentiable horizons}},}\ }\href {\doibase
  10.1007/s002200050336} {\bibfield  {journal} {\bibinfo  {journal} {Commun.
  Math. Phys.}\ }\textbf {\bibinfo {volume} {193}},\ \bibinfo {pages}
  {449--470} (\bibinfo {year} {1998})},\ \Eprint
  {http://arxiv.org/abs/gr-qc/9611032} {arXiv:gr-qc/9611032 [gr-qc]}
  \BibitemShut {NoStop}%
\bibitem [{\citenamefont {Hawking}\ and\ \citenamefont
  {Hartle}(1972)}]{Hawking:1972hy}%
  \BibitemOpen
  \bibfield  {author} {\bibinfo {author} {\bibfnamefont {S.~W.}\ \bibnamefont
  {Hawking}}\ and\ \bibinfo {author} {\bibfnamefont {J.~B.}\ \bibnamefont
  {Hartle}},\ }\bibfield  {title} {\enquote {\bibinfo {title} {{Energy and
  angular momentum flow into a black hole}},}\ }\href {\doibase
  10.1007/BF01645515} {\bibfield  {journal} {\bibinfo  {journal} {Commun. Math.
  Phys.}\ }\textbf {\bibinfo {volume} {27}},\ \bibinfo {pages} {283--290}
  (\bibinfo {year} {1972})}\BibitemShut {NoStop}%
\bibitem [{\citenamefont {Ashtekar}\ and\ \citenamefont
  {Krishnan}(2004)}]{Ashtekar:2004cn}%
  \BibitemOpen
  \bibfield  {author} {\bibinfo {author} {\bibfnamefont {Abhay}\ \bibnamefont
  {Ashtekar}}\ and\ \bibinfo {author} {\bibfnamefont {Badri}\ \bibnamefont
  {Krishnan}},\ }\bibfield  {title} {\enquote {\bibinfo {title} {{Isolated and
  dynamical horizons and their applications}},}\ }\href@noop {} {\bibfield
  {journal} {\bibinfo  {journal} {Living Rev. Rel.}\ }\textbf {\bibinfo
  {volume} {7}},\ \bibinfo {pages} {10} (\bibinfo {year} {2004})},\ \Eprint
  {http://arxiv.org/abs/gr-qc/0407042} {arXiv:gr-qc/0407042} \BibitemShut
  {NoStop}%
\bibitem [{\citenamefont {Penrose}(1965)}]{Penrose:1964wq}%
  \BibitemOpen
  \bibfield  {author} {\bibinfo {author} {\bibfnamefont {Roger}\ \bibnamefont
  {Penrose}},\ }\bibfield  {title} {\enquote {\bibinfo {title} {{Gravitational
  collapse and space-time singularities}},}\ }\href {\doibase
  10.1103/PhysRevLett.14.57} {\bibfield  {journal} {\bibinfo  {journal} {Phys.
  Rev. Lett.}\ }\textbf {\bibinfo {volume} {14}},\ \bibinfo {pages} {57--59}
  (\bibinfo {year} {1965})}\BibitemShut {NoStop}%
\bibitem [{\citenamefont {Booth}(2005)}]{Booth:2005qc}%
  \BibitemOpen
  \bibfield  {author} {\bibinfo {author} {\bibfnamefont {Ivan}\ \bibnamefont
  {Booth}},\ }\bibfield  {title} {\enquote {\bibinfo {title} {{Black hole
  boundaries}},}\ }\href {\doibase 10.1139/p05-063} {\bibfield  {journal}
  {\bibinfo  {journal} {Can. J. Phys.}\ }\textbf {\bibinfo {volume} {83}},\
  \bibinfo {pages} {1073--1099} (\bibinfo {year} {2005})},\ \Eprint
  {http://arxiv.org/abs/gr-qc/0508107} {arXiv:gr-qc/0508107} \BibitemShut
  {NoStop}%
\bibitem [{\citenamefont {Faraoni}\ and\ \citenamefont
  {Prain}(2015)}]{Faraoni:2015pmn}%
  \BibitemOpen
  \bibfield  {author} {\bibinfo {author} {\bibfnamefont {Valerio}\ \bibnamefont
  {Faraoni}}\ and\ \bibinfo {author} {\bibfnamefont {Angus}\ \bibnamefont
  {Prain}},\ }\bibfield  {title} {\enquote {\bibinfo {title} {{Understanding
  dynamical black hole apparent horizons}},}\ }\href@noop {} {\bibfield
  {journal} {\bibinfo  {journal} {Lecture Notes in Physics}\ }\textbf {\bibinfo
  {volume} {907}},\ \bibinfo {pages} {1--199} (\bibinfo {year} {2015})},\
  \Eprint {http://arxiv.org/abs/1511.07775} {arXiv:1511.07775 [gr-qc]}
  \BibitemShut {NoStop}%
\bibitem [{\citenamefont {Krishnan}(2014)}]{Krishnan:2013saa}%
  \BibitemOpen
  \bibfield  {author} {\bibinfo {author} {\bibfnamefont {Badri}\ \bibnamefont
  {Krishnan}},\ }\bibfield  {title} {\enquote {\bibinfo {title} {{Quasi-local
  black hole horizons}},}\ }in\ \href {\doibase 10.1007/978-3-642-41992-8_25}
  {\emph {\bibinfo {booktitle} {Springer Handbook of Spacetime}}},\ \bibinfo
  {editor} {edited by\ \bibinfo {editor} {\bibfnamefont {Abhay}\ \bibnamefont
  {Ashtekar}}\ and\ \bibinfo {editor} {\bibfnamefont {Vesselin}\ \bibnamefont
  {Petkov}}}\ (\bibinfo  {publisher} {Springer-Verlag},\ \bibinfo {year}
  {2014})\ pp.\ \bibinfo {pages} {527--555},\ \Eprint
  {http://arxiv.org/abs/1303.4635} {arXiv:1303.4635 [gr-qc]} \BibitemShut
  {NoStop}%
\bibitem [{\citenamefont {Pook-Kolb}\ \emph
  {et~al.}(2019{\natexlab{a}})\citenamefont {Pook-Kolb}, \citenamefont
  {Birnholtz}, \citenamefont {Krishnan},\ and\ \citenamefont
  {Schnetter}}]{Pook-Kolb:2018igu}%
  \BibitemOpen
  \bibfield  {author} {\bibinfo {author} {\bibfnamefont {Daniel}\ \bibnamefont
  {Pook-Kolb}}, \bibinfo {author} {\bibfnamefont {Ofek}\ \bibnamefont
  {Birnholtz}}, \bibinfo {author} {\bibfnamefont {Badri}\ \bibnamefont
  {Krishnan}}, \ and\ \bibinfo {author} {\bibfnamefont {Erik}\ \bibnamefont
  {Schnetter}},\ }\bibfield  {title} {\enquote {\bibinfo {title} {Existence and
  stability of marginally trapped surfaces in black-hole spacetimes},}\ }\href
  {\doibase 10.1103/PhysRevD.99.064005} {\bibfield  {journal} {\bibinfo
  {journal} {Phys. Rev. D}\ }\textbf {\bibinfo {volume} {99}},\ \bibinfo
  {pages} {064005} (\bibinfo {year} {2019}{\natexlab{a}})}\BibitemShut
  {NoStop}%
\bibitem [{\citenamefont {Pook-Kolb}\ \emph
  {et~al.}(2019{\natexlab{b}})\citenamefont {Pook-Kolb}, \citenamefont
  {Birnholtz}, \citenamefont {Krishnan},\ and\ \citenamefont
  {Schnetter}}]{Pook-Kolb:2019iao}%
  \BibitemOpen
  \bibfield  {author} {\bibinfo {author} {\bibfnamefont {Daniel}\ \bibnamefont
  {Pook-Kolb}}, \bibinfo {author} {\bibfnamefont {Ofek}\ \bibnamefont
  {Birnholtz}}, \bibinfo {author} {\bibfnamefont {Badri}\ \bibnamefont
  {Krishnan}}, \ and\ \bibinfo {author} {\bibfnamefont {Erik}\ \bibnamefont
  {Schnetter}},\ }\bibfield  {title} {\enquote {\bibinfo {title} {{The interior
  of a binary black hole merger}},}\ }\href@noop {} {\  (\bibinfo {year}
  {2019}{\natexlab{b}})},\ \Eprint {http://arxiv.org/abs/1903.05626}
  {arXiv:1903.05626 [gr-qc]} \BibitemShut {NoStop}%
\bibitem [{\citenamefont {Gourgoulhon}\ and\ \citenamefont
  {Jaramillo}(2006)}]{Gourgoulhon:2005ng}%
  \BibitemOpen
  \bibfield  {author} {\bibinfo {author} {\bibfnamefont {Eric}\ \bibnamefont
  {Gourgoulhon}}\ and\ \bibinfo {author} {\bibfnamefont {Jose~Luis}\
  \bibnamefont {Jaramillo}},\ }\bibfield  {title} {\enquote {\bibinfo {title}
  {{A 3+1 perspective on null hypersurfaces and isolated horizons}},}\ }\href
  {\doibase 10.1016/j.physrep.2005.10.005} {\bibfield  {journal} {\bibinfo
  {journal} {Phys. Rept.}\ }\textbf {\bibinfo {volume} {423}},\ \bibinfo
  {pages} {159--294} (\bibinfo {year} {2006})},\ \Eprint
  {http://arxiv.org/abs/gr-qc/0503113} {arXiv:gr-qc/0503113} \BibitemShut
  {NoStop}%
\bibitem [{\citenamefont {Hayward}(2004)}]{Hayward:2004fz}%
  \BibitemOpen
  \bibfield  {author} {\bibinfo {author} {\bibfnamefont {Sean~A.}\ \bibnamefont
  {Hayward}},\ }\bibfield  {title} {\enquote {\bibinfo {title} {{Energy and
  entropy conservation for dynamical black holes}},}\ }\href {\doibase
  10.1103/PhysRevD.70.104027} {\bibfield  {journal} {\bibinfo  {journal} {Phys.
  Rev.}\ }\textbf {\bibinfo {volume} {D70}},\ \bibinfo {pages} {104027}
  (\bibinfo {year} {2004})},\ \Eprint {http://arxiv.org/abs/gr-qc/0408008}
  {arXiv:gr-qc/0408008} \BibitemShut {NoStop}%
\bibitem [{\citenamefont {Jaramillo}(2011)}]{Jaramillo:2011zw}%
  \BibitemOpen
  \bibfield  {author} {\bibinfo {author} {\bibfnamefont {Jose~Luis}\
  \bibnamefont {Jaramillo}},\ }\bibfield  {title} {\enquote {\bibinfo {title}
  {{An introduction to local Black Hole horizons in the 3+1 approach to General
  Relativity}},}\ }\href {\doibase 10.1142/S0218271811020366} {\bibfield
  {journal} {\bibinfo  {journal} {Int. J. Mod. Phys.}\ }\textbf {\bibinfo
  {volume} {D20}},\ \bibinfo {pages} {2169} (\bibinfo {year} {2011})},\ \Eprint
  {http://arxiv.org/abs/1108.2408} {arXiv:1108.2408 [gr-qc]} \BibitemShut
  {NoStop}%
\bibitem [{\citenamefont {Krishnan}(2008)}]{Krishnan:2007va}%
  \BibitemOpen
  \bibfield  {author} {\bibinfo {author} {\bibfnamefont {Badri}\ \bibnamefont
  {Krishnan}},\ }\bibfield  {title} {\enquote {\bibinfo {title} {{Fundamental
  properties and applications of quasi-local black hole horizons}},}\
  }\bibfield  {booktitle} {\emph {\bibinfo {booktitle} {{Proceedings, 18th
  International Conference on General Relativity and Gravitation (GRG18) and
  7th Edoardo Amaldi Conference on Gravitational Waves (Amaldi7), Sydney,
  Australia, July 2007}}},\ }\href {\doibase 10.1088/0264-9381/25/11/114005}
  {\bibfield  {journal} {\bibinfo  {journal} {Class. Quant. Grav.}\ }\textbf
  {\bibinfo {volume} {25}},\ \bibinfo {pages} {114005} (\bibinfo {year}
  {2008})},\ \Eprint {http://arxiv.org/abs/0712.1575} {arXiv:0712.1575 [gr-qc]}
  \BibitemShut {NoStop}%
\bibitem [{\citenamefont {Dreyer}\ \emph {et~al.}(2003)\citenamefont {Dreyer},
  \citenamefont {Krishnan}, \citenamefont {Shoemaker},\ and\ \citenamefont
  {Schnetter}}]{Dreyer:2002mx}%
  \BibitemOpen
  \bibfield  {author} {\bibinfo {author} {\bibfnamefont {Olaf}\ \bibnamefont
  {Dreyer}}, \bibinfo {author} {\bibfnamefont {Badri}\ \bibnamefont
  {Krishnan}}, \bibinfo {author} {\bibfnamefont {Deirdre}\ \bibnamefont
  {Shoemaker}}, \ and\ \bibinfo {author} {\bibfnamefont {Erik}\ \bibnamefont
  {Schnetter}},\ }\bibfield  {title} {\enquote {\bibinfo {title} {{Introduction
  to Isolated Horizons in Numerical Relativity}},}\ }\href {\doibase
  10.1103/PhysRevD.67.024018} {\bibfield  {journal} {\bibinfo  {journal} {Phys.
  Rev.}\ }\textbf {\bibinfo {volume} {D67}},\ \bibinfo {pages} {024018}
  (\bibinfo {year} {2003})},\ \Eprint {http://arxiv.org/abs/gr-qc/0206008}
  {arXiv:gr-qc/0206008} \BibitemShut {NoStop}%
\bibitem [{\citenamefont {Schnetter}\ \emph {et~al.}(2006)\citenamefont
  {Schnetter}, \citenamefont {Krishnan},\ and\ \citenamefont
  {Beyer}}]{Schnetter:2006yt}%
  \BibitemOpen
  \bibfield  {author} {\bibinfo {author} {\bibfnamefont {Erik}\ \bibnamefont
  {Schnetter}}, \bibinfo {author} {\bibfnamefont {Badri}\ \bibnamefont
  {Krishnan}}, \ and\ \bibinfo {author} {\bibfnamefont {Florian}\ \bibnamefont
  {Beyer}},\ }\bibfield  {title} {\enquote {\bibinfo {title} {{Introduction to
  dynamical horizons in numerical relativity}},}\ }\href {\doibase
  10.1103/PhysRevD.74.024028} {\bibfield  {journal} {\bibinfo  {journal} {Phys.
  Rev.}\ }\textbf {\bibinfo {volume} {D74}},\ \bibinfo {pages} {024028}
  (\bibinfo {year} {2006})},\ \Eprint {http://arxiv.org/abs/gr-qc/0604015}
  {arXiv:gr-qc/0604015} \BibitemShut {NoStop}%
\bibitem [{\citenamefont {Newman}(1987)}]{Newman1987}%
  \BibitemOpen
  \bibfield  {author} {\bibinfo {author} {\bibfnamefont {R~P A~C}\ \bibnamefont
  {Newman}},\ }\bibfield  {title} {\enquote {\bibinfo {title} {Topology and
  stability of marginal 2-surfaces},}\ }\href
  {http://stacks.iop.org/0264-9381/4/i=2/a=011} {\bibfield  {journal} {\bibinfo
   {journal} {Classical and Quantum Gravity}\ }\textbf {\bibinfo {volume}
  {4}},\ \bibinfo {pages} {277} (\bibinfo {year} {1987})}\BibitemShut {NoStop}%
\bibitem [{\citenamefont {Andersson}\ \emph {et~al.}(2005)\citenamefont
  {Andersson}, \citenamefont {Mars},\ and\ \citenamefont
  {Simon}}]{Andersson:2005gq}%
  \BibitemOpen
  \bibfield  {author} {\bibinfo {author} {\bibfnamefont {Lars}\ \bibnamefont
  {Andersson}}, \bibinfo {author} {\bibfnamefont {Marc}\ \bibnamefont {Mars}},
  \ and\ \bibinfo {author} {\bibfnamefont {Walter}\ \bibnamefont {Simon}},\
  }\bibfield  {title} {\enquote {\bibinfo {title} {{Local existence of
  dynamical and trapping horizons}},}\ }\href {\doibase
  10.1103/PhysRevLett.95.111102} {\bibfield  {journal} {\bibinfo  {journal}
  {Phys.Rev.Lett.}\ }\textbf {\bibinfo {volume} {95}},\ \bibinfo {pages}
  {111102} (\bibinfo {year} {2005})},\ \Eprint
  {http://arxiv.org/abs/gr-qc/0506013} {arXiv:gr-qc/0506013 [gr-qc]}
  \BibitemShut {NoStop}%
\bibitem [{\citenamefont {Andersson}\ \emph {et~al.}(2008)\citenamefont
  {Andersson}, \citenamefont {Mars},\ and\ \citenamefont
  {Simon}}]{Andersson:2007fh}%
  \BibitemOpen
  \bibfield  {author} {\bibinfo {author} {\bibfnamefont {Lars}\ \bibnamefont
  {Andersson}}, \bibinfo {author} {\bibfnamefont {Marc}\ \bibnamefont {Mars}},
  \ and\ \bibinfo {author} {\bibfnamefont {Walter}\ \bibnamefont {Simon}},\
  }\bibfield  {title} {\enquote {\bibinfo {title} {{Stability of marginally
  outer trapped surfaces and existence of marginally outer trapped tubes}},}\
  }\href@noop {} {\bibfield  {journal} {\bibinfo  {journal}
  {Adv.Theor.Math.Phys.}\ }\textbf {\bibinfo {volume} {12}} (\bibinfo {year}
  {2008})},\ \Eprint {http://arxiv.org/abs/0704.2889} {arXiv:0704.2889 [gr-qc]}
  \BibitemShut {NoStop}%
\bibitem [{\citenamefont {Andersson}\ \emph {et~al.}(2009)\citenamefont
  {Andersson}, \citenamefont {Mars}, \citenamefont {Metzger},\ and\
  \citenamefont {Simon}}]{Andersson:2008up}%
  \BibitemOpen
  \bibfield  {author} {\bibinfo {author} {\bibfnamefont {Lars}\ \bibnamefont
  {Andersson}}, \bibinfo {author} {\bibfnamefont {Marc}\ \bibnamefont {Mars}},
  \bibinfo {author} {\bibfnamefont {Jan}\ \bibnamefont {Metzger}}, \ and\
  \bibinfo {author} {\bibfnamefont {Walter}\ \bibnamefont {Simon}},\ }\bibfield
   {title} {\enquote {\bibinfo {title} {{The Time evolution of marginally
  trapped surfaces}},}\ }\href {\doibase 10.1088/0264-9381/26/8/085018}
  {\bibfield  {journal} {\bibinfo  {journal} {Class.Quant.Grav.}\ }\textbf
  {\bibinfo {volume} {26}},\ \bibinfo {pages} {085018} (\bibinfo {year}
  {2009})},\ \Eprint {http://arxiv.org/abs/0811.4721} {arXiv:0811.4721 [gr-qc]}
  \BibitemShut {NoStop}%
\bibitem [{\citenamefont {Booth}\ \emph {et~al.}(2017)\citenamefont {Booth},
  \citenamefont {Kunduri},\ and\ \citenamefont {O'Grady}}]{Booth:2017fob}%
  \BibitemOpen
  \bibfield  {author} {\bibinfo {author} {\bibfnamefont {Ivan}\ \bibnamefont
  {Booth}}, \bibinfo {author} {\bibfnamefont {Hari~K.}\ \bibnamefont
  {Kunduri}}, \ and\ \bibinfo {author} {\bibfnamefont {Anna}\ \bibnamefont
  {O'Grady}},\ }\bibfield  {title} {\enquote {\bibinfo {title} {{Unstable
  marginally outer trapped surfaces in static spherically symmetric
  spacetimes}},}\ }\href {\doibase 10.1103/PhysRevD.96.024059} {\bibfield
  {journal} {\bibinfo  {journal} {Phys. Rev.}\ }\textbf {\bibinfo {volume}
  {D96}},\ \bibinfo {pages} {024059} (\bibinfo {year} {2017})},\ \Eprint
  {http://arxiv.org/abs/1705.03063} {arXiv:1705.03063 [gr-qc]} \BibitemShut
  {NoStop}%
\bibitem [{\citenamefont {Sherif}\ \emph {et~al.}(2018)\citenamefont {Sherif},
  \citenamefont {Goswami},\ and\ \citenamefont {Maharaj}}]{Sherif:2018scu}%
  \BibitemOpen
  \bibfield  {author} {\bibinfo {author} {\bibfnamefont {Abbas}\ \bibnamefont
  {Sherif}}, \bibinfo {author} {\bibfnamefont {Rituparno}\ \bibnamefont
  {Goswami}}, \ and\ \bibinfo {author} {\bibfnamefont {Sunil~D.}\ \bibnamefont
  {Maharaj}},\ }\bibfield  {title} {\enquote {\bibinfo {title} {{Geometrical
  properties of trapped surfaces and apparent horizons}},}\ }\href@noop {} {\
  (\bibinfo {year} {2018})},\ \Eprint {http://arxiv.org/abs/1805.05684}
  {arXiv:1805.05684 [gr-qc]} \BibitemShut {NoStop}%
\bibitem [{\citenamefont {Mach}\ and\ \citenamefont
  {Xie}(2017)}]{Mach:2017peu}%
  \BibitemOpen
  \bibfield  {author} {\bibinfo {author} {\bibfnamefont {Patryk}\ \bibnamefont
  {Mach}}\ and\ \bibinfo {author} {\bibfnamefont {Naqing}\ \bibnamefont
  {Xie}},\ }\bibfield  {title} {\enquote {\bibinfo {title} {{Toroidal
  marginally outer trapped surfaces in closed
  Friedmann-Lemaître-Robertson-Walker spacetimes: Stability and isoperimetric
  inequalities}},}\ }\href {\doibase 10.1103/PhysRevD.96.084050} {\bibfield
  {journal} {\bibinfo  {journal} {Phys. Rev.}\ }\textbf {\bibinfo {volume}
  {D96}},\ \bibinfo {pages} {084050} (\bibinfo {year} {2017})},\ \Eprint
  {http://arxiv.org/abs/1706.07594} {arXiv:1706.07594 [gr-qc]} \BibitemShut
  {NoStop}%
\bibitem [{\citenamefont {Thornburg}(2004)}]{Thornburg:2003sf}%
  \BibitemOpen
  \bibfield  {author} {\bibinfo {author} {\bibfnamefont {Jonathan}\
  \bibnamefont {Thornburg}},\ }\bibfield  {title} {\enquote {\bibinfo {title}
  {{A Fast Apparent-Horizon Finder for 3-Dimensional Cartesian Grids in
  Numerical Relativity}},}\ }\href {\doibase 10.1088/0264-9381/21/2/026}
  {\bibfield  {journal} {\bibinfo  {journal} {Class. Quant. Grav.}\ }\textbf
  {\bibinfo {volume} {21}},\ \bibinfo {pages} {743--766} (\bibinfo {year}
  {2004})},\ \Eprint {http://arxiv.org/abs/gr-qc/0306056} {arXiv:gr-qc/0306056}
  \BibitemShut {NoStop}%
\bibitem [{\citenamefont {Jaramillo}\ \emph {et~al.}(2009)\citenamefont
  {Jaramillo}, \citenamefont {Ansorg},\ and\ \citenamefont
  {Vasset}}]{Jaramillo:2009zz}%
  \BibitemOpen
  \bibfield  {author} {\bibinfo {author} {\bibfnamefont {Jose~Luis}\
  \bibnamefont {Jaramillo}}, \bibinfo {author} {\bibfnamefont {Marcus}\
  \bibnamefont {Ansorg}}, \ and\ \bibinfo {author} {\bibfnamefont {Nicolas}\
  \bibnamefont {Vasset}},\ }\bibfield  {title} {\enquote {\bibinfo {title}
  {{Application of initial data sequences to the study of black hole dynamical
  trapping horizons}},}\ }\bibfield  {booktitle} {\emph {\bibinfo {booktitle}
  {{Physics and mathematical of gravitation. Proceedings, Spanish Relativity
  Meeting, Salamanca, Spain, September 15-19, 2008}}},\ }\href {\doibase
  10.1063/1.3141305} {\bibfield  {journal} {\bibinfo  {journal} {AIP Conf.
  Proc.}\ }\textbf {\bibinfo {volume} {1122}},\ \bibinfo {pages} {308--311}
  (\bibinfo {year} {2009})},\ \Eprint {http://arxiv.org/abs/1103.6180}
  {arXiv:1103.6180 [gr-qc]} \BibitemShut {NoStop}%
\bibitem [{\citenamefont {Ansorg}(2005)}]{Ansorg:2005bp}%
  \BibitemOpen
  \bibfield  {author} {\bibinfo {author} {\bibfnamefont {Marcus}\ \bibnamefont
  {Ansorg}},\ }\bibfield  {title} {\enquote {\bibinfo {title} {{Double-domain
  spectral method for black hole excision data}},}\ }\href {\doibase
  10.1103/PhysRevD.72.024018} {\bibfield  {journal} {\bibinfo  {journal} {Phys.
  Rev.}\ }\textbf {\bibinfo {volume} {D72}},\ \bibinfo {pages} {024018}
  (\bibinfo {year} {2005})},\ \Eprint {http://arxiv.org/abs/gr-qc/0505059}
  {arXiv:gr-qc/0505059 [gr-qc]} \BibitemShut {NoStop}%
\bibitem [{\citenamefont {Matzner}\ \emph {et~al.}(1999)\citenamefont
  {Matzner}, \citenamefont {Huq},\ and\ \citenamefont
  {Shoemaker}}]{Matzner:1998pt}%
  \BibitemOpen
  \bibfield  {author} {\bibinfo {author} {\bibfnamefont {Richard~A.}\
  \bibnamefont {Matzner}}, \bibinfo {author} {\bibfnamefont {Mijan~F.}\
  \bibnamefont {Huq}}, \ and\ \bibinfo {author} {\bibfnamefont {Deirdre}\
  \bibnamefont {Shoemaker}},\ }\bibfield  {title} {\enquote {\bibinfo {title}
  {{Initial data and coordinates for multiple black hole systems}},}\ }\href
  {\doibase 10.1103/PhysRevD.59.024015} {\bibfield  {journal} {\bibinfo
  {journal} {Phys. Rev.}\ }\textbf {\bibinfo {volume} {D59}},\ \bibinfo {pages}
  {024015} (\bibinfo {year} {1999})},\ \Eprint
  {http://arxiv.org/abs/gr-qc/9805023} {arXiv:gr-qc/9805023 [gr-qc]}
  \BibitemShut {NoStop}%
\bibitem [{\citenamefont {Brill}\ and\ \citenamefont
  {Lindquist}(1963)}]{PhysRev.131.471}%
  \BibitemOpen
  \bibfield  {author} {\bibinfo {author} {\bibfnamefont {Dieter~R.}\
  \bibnamefont {Brill}}\ and\ \bibinfo {author} {\bibfnamefont {Richard~W.}\
  \bibnamefont {Lindquist}},\ }\bibfield  {title} {\enquote {\bibinfo {title}
  {Interaction energy in geometrostatics},}\ }\href {\doibase
  10.1103/PhysRev.131.471} {\bibfield  {journal} {\bibinfo  {journal} {Phys.
  Rev.}\ }\textbf {\bibinfo {volume} {131}},\ \bibinfo {pages} {471--476}
  (\bibinfo {year} {1963})}\BibitemShut {NoStop}%
\bibitem [{\citenamefont {Alcubierre}\ \emph {et~al.}(2000)\citenamefont
  {Alcubierre}, \citenamefont {Allen}, \citenamefont {Br{\"u}gmann},
  \citenamefont {Dramlitsch}, \citenamefont {Font}, \citenamefont
  {Papadopoulos}, \citenamefont {Seidel}, \citenamefont {Stergioulas},
  \citenamefont {Suen},\ and\ \citenamefont {Takahashi}}]{Alcubierre:2000xu}%
  \BibitemOpen
  \bibfield  {author} {\bibinfo {author} {\bibfnamefont {Miguel}\ \bibnamefont
  {Alcubierre}}, \bibinfo {author} {\bibfnamefont {Gabrielle}\ \bibnamefont
  {Allen}}, \bibinfo {author} {\bibfnamefont {Bernd}\ \bibnamefont
  {Br{\"u}gmann}}, \bibinfo {author} {\bibfnamefont {Thomas}\ \bibnamefont
  {Dramlitsch}}, \bibinfo {author} {\bibfnamefont {Jose~A.}\ \bibnamefont
  {Font}}, \bibinfo {author} {\bibfnamefont {Philippos}\ \bibnamefont
  {Papadopoulos}}, \bibinfo {author} {\bibfnamefont {Edward}\ \bibnamefont
  {Seidel}}, \bibinfo {author} {\bibfnamefont {Nikolaos}\ \bibnamefont
  {Stergioulas}}, \bibinfo {author} {\bibfnamefont {Wai-Mo}\ \bibnamefont
  {Suen}}, \ and\ \bibinfo {author} {\bibfnamefont {Ryoji}\ \bibnamefont
  {Takahashi}},\ }\bibfield  {title} {\enquote {\bibinfo {title} {{Towards a
  stable numerical evolution of strongly gravitating systems in general
  relativity: The Conformal treatments}},}\ }\href {\doibase
  10.1103/PhysRevD.62.044034} {\bibfield  {journal} {\bibinfo  {journal} {Phys.
  Rev.}\ }\textbf {\bibinfo {volume} {D62}},\ \bibinfo {pages} {044034}
  (\bibinfo {year} {2000})},\ \Eprint {http://arxiv.org/abs/gr-qc/0003071}
  {arXiv:gr-qc/0003071 [gr-qc]} \BibitemShut {NoStop}%
\bibitem [{\citenamefont {Alcubierre}\ \emph {et~al.}(2003)\citenamefont
  {Alcubierre}, \citenamefont {Br{\"u}gmann}, \citenamefont {Diener},
  \citenamefont {Koppitz}, \citenamefont {Pollney}, \citenamefont {Seidel},\
  and\ \citenamefont {Takahashi}}]{Alcubierre:2002kk}%
  \BibitemOpen
  \bibfield  {author} {\bibinfo {author} {\bibfnamefont {Miguel}\ \bibnamefont
  {Alcubierre}}, \bibinfo {author} {\bibfnamefont {Bernd}\ \bibnamefont
  {Br{\"u}gmann}}, \bibinfo {author} {\bibfnamefont {Peter}\ \bibnamefont
  {Diener}}, \bibinfo {author} {\bibfnamefont {Michael}\ \bibnamefont
  {Koppitz}}, \bibinfo {author} {\bibfnamefont {Denis}\ \bibnamefont
  {Pollney}}, \bibinfo {author} {\bibfnamefont {Edward}\ \bibnamefont
  {Seidel}}, \ and\ \bibinfo {author} {\bibfnamefont {Ryoji}\ \bibnamefont
  {Takahashi}},\ }\bibfield  {title} {\enquote {\bibinfo {title} {{Gauge
  conditions for long term numerical black hole evolutions without
  excision}},}\ }\href {\doibase 10.1103/PhysRevD.67.084023} {\bibfield
  {journal} {\bibinfo  {journal} {Phys. Rev.}\ }\textbf {\bibinfo {volume}
  {D67}},\ \bibinfo {pages} {084023} (\bibinfo {year} {2003})},\ \Eprint
  {http://arxiv.org/abs/gr-qc/0206072} {arXiv:gr-qc/0206072 [gr-qc]}
  \BibitemShut {NoStop}%
\bibitem [{\citenamefont {Pretorius}(2005{\natexlab{b}})}]{Pretorius_2005}%
  \BibitemOpen
  \bibfield  {author} {\bibinfo {author} {\bibfnamefont {Frans}\ \bibnamefont
  {Pretorius}},\ }\bibfield  {title} {\enquote {\bibinfo {title} {Numerical
  relativity using a generalized harmonic decomposition},}\ }\href {\doibase
  10.1088/0264-9381/22/2/014} {\bibfield  {journal} {\bibinfo  {journal}
  {Classical and Quantum Gravity}\ }\textbf {\bibinfo {volume} {22}},\ \bibinfo
  {pages} {425--451} (\bibinfo {year} {2005}{\natexlab{b}})}\BibitemShut
  {NoStop}%
\bibitem [{\citenamefont {Husa}\ \emph {et~al.}(2006)\citenamefont {Husa},
  \citenamefont {Hinder},\ and\ \citenamefont {Lechner}}]{Husa:2004ip}%
  \BibitemOpen
  \bibfield  {author} {\bibinfo {author} {\bibfnamefont {Sascha}\ \bibnamefont
  {Husa}}, \bibinfo {author} {\bibfnamefont {Ian}\ \bibnamefont {Hinder}}, \
  and\ \bibinfo {author} {\bibfnamefont {Christiane}\ \bibnamefont {Lechner}},\
  }\bibfield  {title} {\enquote {\bibinfo {title} {{Kranc: a Mathematica
  application to generate numerical codes for tensorial evolution
  equations}},}\ }\href@noop {} {\bibfield  {journal} {\bibinfo  {journal}
  {Comput. Phys. Commun.}\ }\textbf {\bibinfo {volume} {174}},\ \bibinfo
  {pages} {983--1004} (\bibinfo {year} {2006})},\ \Eprint
  {http://arxiv.org/abs/arXiv:gr-qc/0404023} {arXiv:gr-qc/0404023} \BibitemShut
  {NoStop}%
\bibitem [{Kranc()}]{Kranc:web}%
  \BibitemOpen
  Kranc,\ \href {http://kranccode.org/} {\enquote {\bibinfo {title} {{Kranc}:
  {Kranc} assembles numerical code},}\ }\BibitemShut {NoStop}%
\bibitem [{\citenamefont {Alcubierre}\ \emph {et~al.}(2001)\citenamefont
  {Alcubierre}, \citenamefont {Brandt}, \citenamefont {Bruegmann},
  \citenamefont {Holz}, \citenamefont {Seidel}, \citenamefont {Takahashi},\
  and\ \citenamefont {Thornburg}}]{Alcubierre:1999ab}%
  \BibitemOpen
  \bibfield  {author} {\bibinfo {author} {\bibfnamefont {Miguel}\ \bibnamefont
  {Alcubierre}}, \bibinfo {author} {\bibfnamefont {Steven}\ \bibnamefont
  {Brandt}}, \bibinfo {author} {\bibfnamefont {Bernd}\ \bibnamefont
  {Bruegmann}}, \bibinfo {author} {\bibfnamefont {Daniel}\ \bibnamefont
  {Holz}}, \bibinfo {author} {\bibfnamefont {Edward}\ \bibnamefont {Seidel}},
  \bibinfo {author} {\bibfnamefont {Ryoji}\ \bibnamefont {Takahashi}}, \ and\
  \bibinfo {author} {\bibfnamefont {Jonathan}\ \bibnamefont {Thornburg}},\
  }\bibfield  {title} {\enquote {\bibinfo {title} {{Symmetry without symmetry:
  Numerical simulation of axisymmetric systems using Cartesian grids}},}\
  }\href {\doibase 10.1142/S0218271801000834} {\bibfield  {journal} {\bibinfo
  {journal} {Int. J. Mod. Phys.}\ }\textbf {\bibinfo {volume} {D10}},\ \bibinfo
  {pages} {273--290} (\bibinfo {year} {2001})},\ \Eprint
  {http://arxiv.org/abs/gr-qc/9908012} {arXiv:gr-qc/9908012 [gr-qc]}
  \BibitemShut {NoStop}%
\bibitem [{\citenamefont {Wardell}\ \emph {et~al.}(2016)\citenamefont
  {Wardell}, \citenamefont {Hinder},\ and\ \citenamefont
  {Bentivegna}}]{wardell_barry_2016_155394}%
  \BibitemOpen
  \bibfield  {author} {\bibinfo {author} {\bibfnamefont {Barry}\ \bibnamefont
  {Wardell}}, \bibinfo {author} {\bibfnamefont {Ian}\ \bibnamefont {Hinder}}, \
  and\ \bibinfo {author} {\bibfnamefont {Eloisa}\ \bibnamefont {Bentivegna}},\
  }\href {\doibase 10.5281/zenodo.155394} {\enquote {\bibinfo {title}
  {{Simulation of GW150914 binary black hole merger using the Einstein
  Toolkit}},}\ } (\bibinfo {year} {2016})\BibitemShut {NoStop}%
\bibitem [{\citenamefont {Pook-Kolb}\ \emph
  {et~al.}(2019{\natexlab{c}})\citenamefont {Pook-Kolb}, \citenamefont
  {Birnholtz}, \citenamefont {Krishnan},\ and\ \citenamefont
  {Schnetter}}]{pook_kolb_daniel_2019_3260171}%
  \BibitemOpen
  \bibfield  {author} {\bibinfo {author} {\bibfnamefont {Daniel}\ \bibnamefont
  {Pook-Kolb}}, \bibinfo {author} {\bibfnamefont {Ofek}\ \bibnamefont
  {Birnholtz}}, \bibinfo {author} {\bibfnamefont {Badri}\ \bibnamefont
  {Krishnan}}, \ and\ \bibinfo {author} {\bibfnamefont {Erik}\ \bibnamefont
  {Schnetter}},\ }\href {\doibase 10.5281/zenodo.3260171} {\enquote {\bibinfo
  {title} {{MOTS Finder version 1.2}},}\ } (\bibinfo {year}
  {2019}{\natexlab{c}})\BibitemShut {NoStop}%
\bibitem [{\citenamefont {L{\"{o}}ffler}\ \emph {et~al.}(2012)\citenamefont
  {L{\"{o}}ffler}, \citenamefont {Faber}, \citenamefont {Bentivegna},
  \citenamefont {Bode}, \citenamefont {Diener}, \citenamefont {Haas},
  \citenamefont {Hinder}, \citenamefont {Mundim}, \citenamefont {Ott},
  \citenamefont {Schnetter}, \citenamefont {Allen}, \citenamefont
  {Campanelli},\ and\ \citenamefont {Laguna}}]{Loffler:2011ay}%
  \BibitemOpen
  \bibfield  {author} {\bibinfo {author} {\bibfnamefont {Frank}\ \bibnamefont
  {L{\"{o}}ffler}}, \bibinfo {author} {\bibfnamefont {Joshua}\ \bibnamefont
  {Faber}}, \bibinfo {author} {\bibfnamefont {Eloisa}\ \bibnamefont
  {Bentivegna}}, \bibinfo {author} {\bibfnamefont {Tanja}\ \bibnamefont
  {Bode}}, \bibinfo {author} {\bibfnamefont {Peter}\ \bibnamefont {Diener}},
  \bibinfo {author} {\bibfnamefont {Roland}\ \bibnamefont {Haas}}, \bibinfo
  {author} {\bibfnamefont {Ian}\ \bibnamefont {Hinder}}, \bibinfo {author}
  {\bibfnamefont {Bruno~C.}\ \bibnamefont {Mundim}}, \bibinfo {author}
  {\bibfnamefont {Christian~D.}\ \bibnamefont {Ott}}, \bibinfo {author}
  {\bibfnamefont {Erik}\ \bibnamefont {Schnetter}}, \bibinfo {author}
  {\bibfnamefont {Gabrielle}\ \bibnamefont {Allen}}, \bibinfo {author}
  {\bibfnamefont {Manuela}\ \bibnamefont {Campanelli}}, \ and\ \bibinfo
  {author} {\bibfnamefont {Pablo}\ \bibnamefont {Laguna}},\ }\bibfield  {title}
  {\enquote {\bibinfo {title} {{{T}he {E}instein {T}oolkit: {A} {C}ommunity
  {C}omputational {I}nfrastructure for {R}elativistic {A}strophysics}},}\
  }\href {\doibase doi:10.1088/0264-9381/29/11/115001} {\bibfield  {journal}
  {\bibinfo  {journal} {Class. Quantum Grav.}\ }\textbf {\bibinfo {volume}
  {29}},\ \bibinfo {pages} {115001} (\bibinfo {year} {2012})},\ \Eprint
  {http://arxiv.org/abs/arXiv:1111.3344 [gr-qc]} {arXiv:1111.3344 [gr-qc]}
  \BibitemShut {NoStop}%
\bibitem [{EinsteinToolkit()}]{EinsteinToolkit:web}%
  \BibitemOpen
  EinsteinToolkit,\ \href@noop {} {\enquote {\bibinfo {title} {{Einstein
  Toolkit}: Open software for relativistic astrophysics},}\ }\bibinfo {note}
  {\url{http://einsteintoolkit.org/}}\BibitemShut {NoStop}%
\bibitem [{\citenamefont {Ansorg}\ \emph {et~al.}(2004)\citenamefont {Ansorg},
  \citenamefont {Br{\"u}gmann},\ and\ \citenamefont {Tichy}}]{Ansorg:2004ds}%
  \BibitemOpen
  \bibfield  {author} {\bibinfo {author} {\bibfnamefont {Marcus}\ \bibnamefont
  {Ansorg}}, \bibinfo {author} {\bibfnamefont {Bernd}\ \bibnamefont
  {Br{\"u}gmann}}, \ and\ \bibinfo {author} {\bibfnamefont {Wolfgang}\
  \bibnamefont {Tichy}},\ }\bibfield  {title} {\enquote {\bibinfo {title} {A
  single-domain spectral method for black hole puncture data},}\ }\href
  {\doibase 10.1103/PhysRevD.70.064011} {\bibfield  {journal} {\bibinfo
  {journal} {Phys. Rev. D}\ }\textbf {\bibinfo {volume} {70}},\ \bibinfo
  {pages} {064011} (\bibinfo {year} {2004})},\ \Eprint
  {http://arxiv.org/abs/arXiv:gr-qc/0404056} {arXiv:gr-qc/0404056} \BibitemShut
  {NoStop}%
\bibitem [{\citenamefont {Brown}\ \emph {et~al.}(2009)\citenamefont {Brown},
  \citenamefont {Diener}, \citenamefont {Sarbach}, \citenamefont {Schnetter},\
  and\ \citenamefont {Tiglio}}]{Brown:2008sb}%
  \BibitemOpen
  \bibfield  {author} {\bibinfo {author} {\bibfnamefont {J.~David}\
  \bibnamefont {Brown}}, \bibinfo {author} {\bibfnamefont {Peter}\ \bibnamefont
  {Diener}}, \bibinfo {author} {\bibfnamefont {Olivier}\ \bibnamefont
  {Sarbach}}, \bibinfo {author} {\bibfnamefont {Erik}\ \bibnamefont
  {Schnetter}}, \ and\ \bibinfo {author} {\bibfnamefont {Manuel}\ \bibnamefont
  {Tiglio}},\ }\bibfield  {title} {\enquote {\bibinfo {title} {{Turduckening
  black holes: an analytical and computational study}},}\ }\href {\doibase
  10.1103/PhysRevD.79.044023} {\bibfield  {journal} {\bibinfo  {journal} {Phys.
  Rev. D}\ }\textbf {\bibinfo {volume} {79}},\ \bibinfo {pages} {044023}
  (\bibinfo {year} {2009})},\ \Eprint {http://arxiv.org/abs/arXiv:0809.3533
  [gr-qc]} {arXiv:0809.3533 [gr-qc]} \BibitemShut {NoStop}%
\bibitem [{\citenamefont {{Whitney, Hassler}}({1937})}]{CM_1937__4__276_0}%
  \BibitemOpen
  \bibfield  {author} {\bibinfo {author} {\bibnamefont {{Whitney, Hassler}}},\
  }\bibfield  {title} {\enquote {\bibinfo {title} {{On regular closed curves in
  the plane}},}\ }\href {{http://www.numdam.org/item/CM_1937__4__276_0}}
  {\bibfield  {journal} {\bibinfo  {journal} {{Compositio Mathematica}}\
  }\textbf {\bibinfo {volume} {{4}}},\ \bibinfo {pages} {{276--284}} (\bibinfo
  {year} {{1937}})}\BibitemShut {NoStop}%
\bibitem [{\citenamefont {Smale}(1959)}]{10.2307/1993205}%
  \BibitemOpen
  \bibfield  {author} {\bibinfo {author} {\bibfnamefont {Stephen}\ \bibnamefont
  {Smale}},\ }\bibfield  {title} {\enquote {\bibinfo {title} {A classification
  of immersions of the two-sphere},}\ }\href
  {http://www.jstor.org/stable/1993205} {\bibfield  {journal} {\bibinfo
  {journal} {Transactions of the American Mathematical Society}\ }\textbf
  {\bibinfo {volume} {90}},\ \bibinfo {pages} {281--290} (\bibinfo {year}
  {1959})}\BibitemShut {NoStop}%
\bibitem [{\citenamefont {Schnetter}\ and\ \citenamefont
  {Miller}(2019)}]{erik_schnetter_2019_3258858}%
  \BibitemOpen
  \bibfield  {author} {\bibinfo {author} {\bibfnamefont {Erik}\ \bibnamefont
  {Schnetter}}\ and\ \bibinfo {author} {\bibfnamefont {Jonah}\ \bibnamefont
  {Miller}},\ }\href {\doibase 10.5281/zenodo.3258858} {\enquote {\bibinfo
  {title} {{eschnett/SimulationIO: New release to trigger Zenodo}},}\ }
  (\bibinfo {year} {2019})\BibitemShut {NoStop}%
\bibitem [{\citenamefont {Jones}\ \emph {et~al.}(2001--)\citenamefont {Jones},
  \citenamefont {Oliphant}, \citenamefont {Peterson} \emph
  {et~al.}}]{Jones_SciPy}%
  \BibitemOpen
  \bibfield  {author} {\bibinfo {author} {\bibfnamefont {Eric}\ \bibnamefont
  {Jones}}, \bibinfo {author} {\bibfnamefont {Travis}\ \bibnamefont
  {Oliphant}}, \bibinfo {author} {\bibfnamefont {Pearu}\ \bibnamefont
  {Peterson}},  \emph {et~al.},\ }\href {http://www.scipy.org/} {\enquote
  {\bibinfo {title} {{SciPy}: Open source scientific tools for {Python}},}\ }
  (\bibinfo {year} {2001--}),\ \bibinfo {note} {[Online; accessed
  <today>]}\BibitemShut {NoStop}%
\bibitem [{\citenamefont {{van der Walt}}\ \emph {et~al.}(2011)\citenamefont
  {{van der Walt}}, \citenamefont {{Colbert}},\ and\ \citenamefont
  {{Varoquaux}}}]{van_der_Walt_NumPy}%
  \BibitemOpen
  \bibfield  {author} {\bibinfo {author} {\bibfnamefont {S.}~\bibnamefont {{van
  der Walt}}}, \bibinfo {author} {\bibfnamefont {S.~C.}\ \bibnamefont
  {{Colbert}}}, \ and\ \bibinfo {author} {\bibfnamefont {G.}~\bibnamefont
  {{Varoquaux}}},\ }\bibfield  {title} {\enquote {\bibinfo {title} {The numpy
  array: A structure for efficient numerical computation},}\ }\href {\doibase
  10.1109/MCSE.2011.37} {\bibfield  {journal} {\bibinfo  {journal} {Computing
  in Science Engineering}\ }\textbf {\bibinfo {volume} {13}},\ \bibinfo {pages}
  {22--30} (\bibinfo {year} {2011})}\BibitemShut {NoStop}%
\bibitem [{\citenamefont {Johansson}\ \emph {et~al.}(2018)\citenamefont
  {Johansson} \emph {et~al.}}]{mpmath}%
  \BibitemOpen
  \bibfield  {author} {\bibinfo {author} {\bibfnamefont {Fredrik}\ \bibnamefont
  {Johansson}} \emph {et~al.},\ }\href@noop {} {\emph {\bibinfo {title}
  {mpmath: a {P}ython library for arbitrary-precision floating-point arithmetic
  (version 1.0.0)}}} (\bibinfo {year} {2018}),\ \bibinfo {note} {{\tt
  http://mpmath.org/}}\BibitemShut {NoStop}%
\bibitem [{\citenamefont {Meurer}\ \emph {et~al.}(2017)\citenamefont {Meurer},
  \citenamefont {Smith}, \citenamefont {Paprocki}, \citenamefont
  {{\v{C}}ert{\'\i}k}, \citenamefont {Kirpichev}, \citenamefont {Rocklin},
  \citenamefont {Kumar}, \citenamefont {Ivanov}, \citenamefont {Moore},
  \citenamefont {Singh} \emph {et~al.}}]{meurer2017sympy}%
  \BibitemOpen
  \bibfield  {author} {\bibinfo {author} {\bibfnamefont {Aaron}\ \bibnamefont
  {Meurer}}, \bibinfo {author} {\bibfnamefont {Christopher~P}\ \bibnamefont
  {Smith}}, \bibinfo {author} {\bibfnamefont {Mateusz}\ \bibnamefont
  {Paprocki}}, \bibinfo {author} {\bibfnamefont {Ond{\v{r}}ej}\ \bibnamefont
  {{\v{C}}ert{\'\i}k}}, \bibinfo {author} {\bibfnamefont {Sergey~B}\
  \bibnamefont {Kirpichev}}, \bibinfo {author} {\bibfnamefont {Matthew}\
  \bibnamefont {Rocklin}}, \bibinfo {author} {\bibfnamefont {AMiT}\
  \bibnamefont {Kumar}}, \bibinfo {author} {\bibfnamefont {Sergiu}\
  \bibnamefont {Ivanov}}, \bibinfo {author} {\bibfnamefont {Jason~K}\
  \bibnamefont {Moore}}, \bibinfo {author} {\bibfnamefont {Sartaj}\
  \bibnamefont {Singh}},  \emph {et~al.},\ }\bibfield  {title} {\enquote
  {\bibinfo {title} {{SymPy}: symbolic computing in {Python}},}\ }\href
  {\doibase 10.7717/peerj-cs.103} {\bibfield  {journal} {\bibinfo  {journal}
  {PeerJ Computer Science}\ }\textbf {\bibinfo {volume} {3}},\ \bibinfo {pages}
  {e103} (\bibinfo {year} {2017})}\BibitemShut {NoStop}%
\bibitem [{\citenamefont {Hunter}(2007)}]{Hunter:2007}%
  \BibitemOpen
  \bibfield  {author} {\bibinfo {author} {\bibfnamefont {J.~D.}\ \bibnamefont
  {Hunter}},\ }\bibfield  {title} {\enquote {\bibinfo {title} {Matplotlib: A 2d
  graphics environment},}\ }\href {\doibase 10.1109/MCSE.2007.55} {\bibfield
  {journal} {\bibinfo  {journal} {Computing in Science \& Engineering}\
  }\textbf {\bibinfo {volume} {9}},\ \bibinfo {pages} {90--95} (\bibinfo {year}
  {2007})}\BibitemShut {NoStop}%
\bibitem [{\citenamefont {Droettboom}\ \emph {et~al.}(2018)\citenamefont
  {Droettboom}, \citenamefont {Caswell}, \citenamefont {Hunter}, \citenamefont
  {Firing}, \citenamefont {Nielsen}, \citenamefont {Lee}, \citenamefont
  {de~Andrade}, \citenamefont {Varoquaux}, \citenamefont {Stansby},
  \citenamefont {Root}, \citenamefont {Elson}, \citenamefont {Dale},
  \citenamefont {Lee}, \citenamefont {May}, \citenamefont {Sepp\"anen},
  \citenamefont {Klymak}, \citenamefont {McDougall}, \citenamefont {Straw},
  \citenamefont {Hobson}, \citenamefont {cgohlke}, \citenamefont {Yu},
  \citenamefont {Ma}, \citenamefont {Vincent}, \citenamefont {Silvester},
  \citenamefont {Moad}, \citenamefont {Katins}, \citenamefont {Kniazev},
  \citenamefont {Hoffmann}, \citenamefont {Ariza},\ and\ \citenamefont
  {W\"urtz}}]{michael_droettboom_2018_1202077}%
  \BibitemOpen
  \bibfield  {author} {\bibinfo {author} {\bibfnamefont {Michael}\ \bibnamefont
  {Droettboom}}, \bibinfo {author} {\bibfnamefont {Thomas~A}\ \bibnamefont
  {Caswell}}, \bibinfo {author} {\bibfnamefont {John}\ \bibnamefont {Hunter}},
  \bibinfo {author} {\bibfnamefont {Eric}\ \bibnamefont {Firing}}, \bibinfo
  {author} {\bibfnamefont {Jens~Hedegaard}\ \bibnamefont {Nielsen}}, \bibinfo
  {author} {\bibfnamefont {Antony}\ \bibnamefont {Lee}}, \bibinfo {author}
  {\bibfnamefont {Elliott~Sales}\ \bibnamefont {de~Andrade}}, \bibinfo {author}
  {\bibfnamefont {Nelle}\ \bibnamefont {Varoquaux}}, \bibinfo {author}
  {\bibfnamefont {David}\ \bibnamefont {Stansby}}, \bibinfo {author}
  {\bibfnamefont {Benjamin}\ \bibnamefont {Root}}, \bibinfo {author}
  {\bibfnamefont {Phil}\ \bibnamefont {Elson}}, \bibinfo {author}
  {\bibfnamefont {Darren}\ \bibnamefont {Dale}}, \bibinfo {author}
  {\bibfnamefont {Jae-Joon}\ \bibnamefont {Lee}}, \bibinfo {author}
  {\bibfnamefont {Ryan}\ \bibnamefont {May}}, \bibinfo {author} {\bibfnamefont
  {Jouni~K.}\ \bibnamefont {Sepp\"anen}}, \bibinfo {author} {\bibfnamefont
  {Jody}\ \bibnamefont {Klymak}}, \bibinfo {author} {\bibfnamefont {Damon}\
  \bibnamefont {McDougall}}, \bibinfo {author} {\bibfnamefont {Andrew}\
  \bibnamefont {Straw}}, \bibinfo {author} {\bibfnamefont {Paul}\ \bibnamefont
  {Hobson}}, \bibinfo {author} {\bibnamefont {cgohlke}}, \bibinfo {author}
  {\bibfnamefont {Tony~S}\ \bibnamefont {Yu}}, \bibinfo {author} {\bibfnamefont
  {Eric}\ \bibnamefont {Ma}}, \bibinfo {author} {\bibfnamefont {Adrien~F.}\
  \bibnamefont {Vincent}}, \bibinfo {author} {\bibfnamefont {Steven}\
  \bibnamefont {Silvester}}, \bibinfo {author} {\bibfnamefont {Charlie}\
  \bibnamefont {Moad}}, \bibinfo {author} {\bibfnamefont {Jan}\ \bibnamefont
  {Katins}}, \bibinfo {author} {\bibfnamefont {Nikita}\ \bibnamefont
  {Kniazev}}, \bibinfo {author} {\bibfnamefont {Tim}\ \bibnamefont {Hoffmann}},
  \bibinfo {author} {\bibfnamefont {Federico}\ \bibnamefont {Ariza}}, \ and\
  \bibinfo {author} {\bibfnamefont {Peter}\ \bibnamefont {W\"urtz}},\ }\href
  {\doibase 10.5281/zenodo.1202077} {\enquote {\bibinfo {title}
  {matplotlib/matplotlib v2.2.2},}\ } (\bibinfo {year} {2018})\BibitemShut
  {NoStop}%
\bibitem [{\citenamefont {Loken}\ \emph {et~al.}(2010)\citenamefont {Loken},
  \citenamefont {Gruner}, \citenamefont {Groer}, \citenamefont {Peltier},
  \citenamefont {Bunn}, \citenamefont {Craig}, \citenamefont {Henriques},
  \citenamefont {Dempsey}, \citenamefont {Yu}, \citenamefont {Chen},
  \citenamefont {Dursi}, \citenamefont {Chong}, \citenamefont {Northrup},
  \citenamefont {Pinto}, \citenamefont {Knecht},\ and\ \citenamefont
  {Zon}}]{Loken_2010}%
  \BibitemOpen
  \bibfield  {author} {\bibinfo {author} {\bibfnamefont {Chris}\ \bibnamefont
  {Loken}}, \bibinfo {author} {\bibfnamefont {Daniel}\ \bibnamefont {Gruner}},
  \bibinfo {author} {\bibfnamefont {Leslie}\ \bibnamefont {Groer}}, \bibinfo
  {author} {\bibfnamefont {Richard}\ \bibnamefont {Peltier}}, \bibinfo {author}
  {\bibfnamefont {Neil}\ \bibnamefont {Bunn}}, \bibinfo {author} {\bibfnamefont
  {Michael}\ \bibnamefont {Craig}}, \bibinfo {author} {\bibfnamefont {Teresa}\
  \bibnamefont {Henriques}}, \bibinfo {author} {\bibfnamefont {Jillian}\
  \bibnamefont {Dempsey}}, \bibinfo {author} {\bibfnamefont {Ching-Hsing}\
  \bibnamefont {Yu}}, \bibinfo {author} {\bibfnamefont {Joseph}\ \bibnamefont
  {Chen}}, \bibinfo {author} {\bibfnamefont {L~Jonathan}\ \bibnamefont
  {Dursi}}, \bibinfo {author} {\bibfnamefont {Jason}\ \bibnamefont {Chong}},
  \bibinfo {author} {\bibfnamefont {Scott}\ \bibnamefont {Northrup}}, \bibinfo
  {author} {\bibfnamefont {Jaime}\ \bibnamefont {Pinto}}, \bibinfo {author}
  {\bibfnamefont {Neil}\ \bibnamefont {Knecht}}, \ and\ \bibinfo {author}
  {\bibfnamefont {Ramses~Van}\ \bibnamefont {Zon}},\ }\bibfield  {title}
  {\enquote {\bibinfo {title} {{SciNet}: Lessons learned from building a
  power-efficient top-20 system and data centre},}\ }\href {\doibase
  10.1088/1742-6596/256/1/012026} {\bibfield  {journal} {\bibinfo  {journal}
  {Journal of Physics: Conference Series}\ }\textbf {\bibinfo {volume} {256}},\
  \bibinfo {pages} {012026} (\bibinfo {year} {2010})}\BibitemShut {NoStop}%
\end{thebibliography}%

\end{document}